\documentclass{aa}  

\usepackage{graphicx}
\usepackage{txfonts}
\usepackage[hidelinks, colorlinks=true,linkcolor=blue,citecolor=blue]{hyperref}

\usepackage{siunitx}[=v2]
\DeclareSIUnit\years{yr}
\DeclareSIUnit\year{yr}
\DeclareSIUnit\dyne{dyn}  %

\usepackage{amsmath}
\usepackage[capitalise]{cleveref} %
\Crefname{section}{Section}{Sections}
\crefname{section}{Sect.}{Sects.}

\usepackage[normalem]{ulem} %
\usepackage{xcolor} %
\usepackage[export]{adjustbox}
\usepackage{amssymb}
\usepackage{wasysym}

\usepackage{comment}
\usepackage{tikz}

\usepackage{orcidlink}

\newcommand*\mycirc[1]{%
  \tikz[baseline=(C.base)]{%
    \node[draw,circle,
          inner sep=0.6pt,
          minimum size=1.05em,
          line width=0.4pt] (C) {
      \vphantom{abcdefgh}%
      \makebox[0pt][c]{%
        \ifx#1b\raisebox{-0.22ex}{#1}\else
        \ifx#1d\raisebox{-0.18ex}{#1}\else
        \ifx#1f\raisebox{-0.20ex}{#1}\else
        \ifx#1g\raisebox{+0.10ex}{#1}\else
        \ifx#1h\raisebox{-0.15ex}{#1}\else
          #1%
        \fi\fi\fi\fi\fi
      }%
    };%
  }\xspace
}

\newcommand{\marker}[1]{(\textbf{#1})}

\usepackage{tabularx}

\renewcommand{\arraystretch}{1.3}
\NewDocumentCommand{\mathleftmoon}{}{{\text{\normalfont\leftmoon}}}

\def\kB{\ensuremath{k_\mathrm{B}}\xspace}
\def\amu{\ensuremath{m_\mathrm{u}}\xspace}

\newcommand{\Mach}{\operatorname{\ensuremath{M\kern-.06em a}\xspace}}
\newcommand{\Msol}{M_\mathrm{\odot}}
\newcommand{\Msun}{M_\mathrm{\odot}}
\newcommand{\Mearth}{M_\mathrm{\oplus}}
\newcommand{\Mpla}{\ensuremath{M_\mathrm{p}}\xspace}
\newcommand{\Rpla}{\ensuremath{R_\mathrm{p}}\xspace}
\newcommand{\Rjup}{R_\mathrm{J}}
\newcommand{\qshock}{Q^+_\mathrm{shock}}
\newcommand{\Laccmax}{\ensuremath{L_{\textnormal{acc,\,max}}}\xspace}
\newcommand{\Mdotgasmax}{\ensuremath{\dot{M}_{\textnormal{gas,\,max}}}\xspace}
\newcommand{\Mjup}{\ensuremath{M_{\mathrm{J}}}\xspace}
\newcommand{\Ljup}{\ensuremath{L_{\mathrm{J}}}\xspace}
\newcommand{\Lsol}{\ensuremath{L_{\mathrm{\odot}}}\xspace}
\newcommand{\Rout}{\ensuremath{R_{\textnormal{out}}}\xspace}
\newcommand{\RBondi}{\ensuremath{R_{\textnormal{B}}}\xspace}
\newcommand{\Ltot}{\ensuremath{L_{\textnormal{tot}}}\xspace}
\newcommand{\Lint}{\ensuremath{L_{\textnormal{int}}}\xspace}
\newcommand{\Lbol}{\ensuremath{L_{\textnormal{bol}}}\xspace}
\newcommand{\Qshock}{Q^+_\mathrm{shock}}

\newcommand{\fpg}{\ensuremath{f_{\textnormal{d/g}}}\xspace}

\newcommand{\fred}{\ensuremath{f_{\textnormal{red}}}\xspace}
\newcommand{\Tsh}{\ensuremath{T_{\textnormal{sh}}}\xspace}
\newcommand{\Teff}{\ensuremath{T_{\textnormal{eff}}}\xspace}
\newcommand{\Tsurf}{\ensuremath{T_{\textnormal{s}}}\xspace}

\def\LT{$L$--$T$\xspace}

\makeatletter
\let\jnl@style=\rm
\def\ref@jnl#1{{\jnl@style#1}}

\def\ptp{\ref@jnl{Prog.~Th.~Phys\@}}   %
\def\rprphys{\ref@jnl{Rep.~Prog.~Phys\@}}   %
\def\natas{\ref@jnl{NatAs}}           %
\def\amjm{\ref@jnl{AmJM}}             %

\makeatother

\makeatletter
\renewcommand*\aa@pageof{, page \thepage{} of \pageref*{LastPage}}
\makeatother

\begin{document} 

   \title{Planetary formation tracks on the Hertzsprung--Russell diagram}

   \subtitle{Visualising the processes of giant planet growth}

   \author{%
    Benedikt Gottstein\,\orcidlink{0009-0001-6180-4163}\inst{\ref{bernWP}}
    \and
    Gabriel-Dominique Marleau\,\orcidlink{0000-0002-2919-7500}\inst{\ref{bernWP},\ref{duisburg},\ref{mpia}}
    \and
    Christoph Mordasini\,\orcidlink{0000-0002-1013-2811}\inst{\ref{bernWP},\ref{bernCSH}}%
    }


    \institute{%
    Division of Space Research \&\ Planetary Sciences, Physics Institute, University of Bern, Sidlerstr.~5, 3012 Bern, Switzerland\\
    \email{benedikt.gottstein@unibe.ch}
      \label{bernWP}
    \and
    Fakult\"at f\"ur Physik, Universit\"at Duisburg--Essen, Lotharstr.~1, 47057 Duisburg, Germany
      \label{duisburg}
    \and
    Max-Planck-Institut f\"ur Astronomie, K\"onigstuhl 17, 69117 Heidelberg, Germany
      \label{mpia}
    \and
    Center for Space and Habitability, University of Bern, Gesellschaftsstr.~6, Bern, 3012, Switzerland
      \label{bernCSH}
    }

   \date{Received in 2026; accepted --}

    \abstract
   {The Hertzsprung--Russell Diagram (HRD) has been a cornerstone of astrophysics, illustrating the relationships between stars' luminosity and effective temperature. Although HRDs have been instrumental in understanding stellar evolution, they have not been often applied to planetary formation.}
   {We extend the HRD framework to visualise planet formation, offering a novel perspective on the physical processes involved. Specifically, we investigate how gas and solid accretion, cooling and contraction, and orbital migration shape these tracks under different formation scenarios.}
   {We use the Bern model to calculate the interior structure of planets during their entire formation and evolution. Additionally, we couple this for the first time to the results of radiation-hydrodynamical simulations to calculate dynamically the accretion shock heating efficiency, providing insight into the cold- and hot-start ambiguity.}
   {The planetary HRDs show three branches, each representing a distinct planet formation phase. The first phase, the ascending branch, is heavily influenced by the size of the smallest accreted bodies and the associated solid accretion rate and also orbital migration. HR tracks in this phase are directed steeply upward. For in-situ planetesimal accretion, one finds analytically $L\propto T^8$. The second phase, the planetary horizontal branch, starts when the gas accretion rate becomes disk-limited and the planet detaches from the disk, subsequently contracting rapidly. The planets move nearly horizontally to the left, with hot accretion, more massive planets, and a pebble accretion scenario leading to an upward bending. The planetary interiors are nearly an ideal gas at detachment, but the increasing electron degeneracy lets the central temperatures decrease, stabilising the radii and halting the rapid movement to the left. The third phase, the descending branch, starts when the effective temperature begins to decrease. In this phase, gas accretion stops and planets join classical constant-mass cooling tracks. The radius changes only weakly, making these tracks go diagonally downwards, with $L{\sim}T^4$.
   Also, there is substantial similarity between our analysed tracks, data from synthetic populations, and observational data.} 
  {Planetary HRDs offer a unique way to link young planets' luminosities and effective temperatures to their formation physics, with pebble and planetesimal accretion producing distinct early-time tracks. Comparisons to currently directly-imaged planets are broadly consistent with our tracks, since these objects are mostly non-accreting or near the end of formation and thus expected to follow standard post-formation cooling evolution. However, observationally populating the short-lived early horizontal branch or ascending branch will be challenging, and interpreting embedded, actively accreting planets will require models that include accretion-shock emission and circumplanetary-disk reprocessing.
  }
  

\titlerunning{Planet formation visualised on the Hertzsprung--Russell diagram}
\authorrunning{Gottstein et al.}

   \maketitle

\section{Introduction}
\label{sec:Introduction}

The Hertzsprung--Russell Diagram \citetext{HRD, after \citealt{hertzsprung_uber_1909} and \citealt{russell_relations_1914}} is a fundamental tool in astrophysics, traditionally used to illustrate the relationships between the luminosity of a star and its surface temperature\footnote{
In precise terms, the original (observational) HRD represents the relationship between magnitude and spectral type. However, for the purposes of this paper, the term HRD is used interchangeably with the (theoretical) luminosity--temperature (LT) diagram.
}. This diagram has been instrumental in understanding stellar evolution, mapping the life cycle of stars from their formation to their final stages. It provides insights into key physical processes, including nuclear fusion, interior structure, and energy transport, making it an indispensable framework in the field. However, while the HRD has been extensively used for studying stars, its application in the context of giant planet formation remains largely unexplored.

This paper aims to bridge this gap by employing the HRD to explain and illustrate the formation process from another perspective. Our concept of the planetary HRD is intended to serve as a step towards a basic framework for future classifications of observed young and forming planets. Until recently, there was no direct observation of the planetary formation process nor of these objects. However, a new era of observations has begun lately in which this is -- although difficult -- possible. Such detections have opened avenues to explore the formation of planets in unprecedented detail and in a direct way. These observations come in three different types: kinematic detections \citep{pinte_kinematic_2019,pinte_nine_2020}, potential gap planets \citep{zhang_disk_2018, lodato_newborn_2019, asensio-torres_SPHERE_2021, ruzza_dbnets2.0_2025}, and directly imaged planets. For the last, the first unambiguous direct discovery of two forming planets was made with PDS70~b \citep{keppler_discovery_2018} and PDS70~c \citep{haffert_two_2019}, and more recently WISPIT~2~b \citep{vanCapelleveen_WISPIT_2025, Close_WISPIT_2025} and V2376~Ori~b \citep{vila_observation_2025} were added to the short list. The planets are detected at several wavelengths, revealing not only the photosphere \citep{muller_orbital_2018} but also the gas accretion shock \citep{wagner_magellan_2018,hashimoto_accretion_2020,zhou_hubblePDS70b_2021} and the circumplanetary disk \citep{isella_detection_2019,christiaens_evidence_2019,benisty_CPD-PDS70c_2021,shibaike_constraints-PDS70bc_2024}. These multi-wavelength observations are pivotal for testing and refining theoretical models. Expanding on this topic is essential also for current instruments like SPHERE or ALMA, or future ones like RISTRETTO (\citealp{lovis_RISTRETTO_2022, lovis_RISTRETTO_2024}; Blackman et al., subm.), the SPHERE+ project \citep{boccaletti_sphere_2020,mazoyer_upgrading_2024}, and several instruments at the Extremely Large Telescope (ELT), for example the Mid-infrared ELT Imager and Spectrograph (METIS; e.g., \citealp{oberg_observing_2023,feldt_high_2024,takami_eltmetis_2025,marleau_hydrogenline_2026}).

This work is a first step in taking up this challenge by investigating on the theoretical side the bolometric luminosity and surface temperature evolution of planets during their full formation and evolution process.
We use a global model that does this in a self-consistently coupled way, the Bern Model (see review e.g.\ in \citealt{burn_handbookPopSynth_2024}), extended by a number of new elements. While clearly an approximation, our approach with a low-dimensional 1D model is not facing the limitations experienced by 2D or 3D hydrodynamic simulations, which currently can run only selected cases for a short time. Our approach allows us to track long-term evolutionary processes and derive population-level insights, similar to how the stellar HRD makes it possible to track certain phases of the life of a star and ultimately link them to its interior physics.

In the present paper, we focus on the forming planet itself -- by definition of the HRD -- and assume spherically symmetric accretion. Other aspects such as a potential circumplanetary disc, non-spherical gas accretion or the effects of the surrounding protoplanetary disc and infalling gas on the observability will be addressed in subsequent works.

This paper is structured as follows.
We review in \cref{sec:Model} the main approach for the calculation of the luminosity $L$ and the surface temperature $T$. In this work, we have extended the model to address the cold-hot-start ambiguity arising from gas accretion at the accretion shock using the results of \citet{marleau_accretionShock-II-2019}, and describe this in \cref{sec:shock_conditions}. It yields the fraction of accretion luminosity that is radiated away or alternatively incorporated into the growing planet.
In \cref{sec:Methods+main+var}, we describe our nominal initial conditions, resulting in an HRD of our so-called default case, and then compare and discuss variations in the setup to help understand different processes or explore new parameter spaces.
In particular, in \Cref{sec:twoPopComparison} then we introduce a new formation scenario with the addition of pebbles and the two-population model of \citet{birnstiel_simple_2012}. Through these comparisons, we aim to identify trends and features that could become the foundation for interpreting future observations.
Finally, in \cref{sec:summary} we summarise our findings before concluding in \cref{sec:discussion} and comparing our tracks with data from synthetic populations and observations.

\section{Model}
\label{sec:Model}
The Bern model is described in detail in \cite{emsenhuber_ngpps1_2021}.
In short, the Bern model is a global planet formation and evolution model which includes the calculation of the 1D spherically interior structure of (proto)planets, and which is designed to study the formation and long-term evolution of planetary systems. In this work, however, we will focus on single-embryo formation scenarios. 
In this section we concentrate on relevant model updates and key points regarding the calculation of the luminosity and surface temperature, which is based on the work of \citet{mordasini_characterization-I_2012}.
\subsection{Envelope structure and luminosity calculation}
\label{sec:envelopeStructureUpdated}

Compared with \citet{mordasini_characterization-I_2012}, the basic 1D envelope structure equations remain unchanged, so we do not repeat them here (see the original work for the derivation and numerical implementation). The only update relevant for the present paper is the opacity treatment, where we combine radiative and conductive opacities harmonically, effectively approximately taking their minimum:
\begin{align}
\label{eq:opacityRad}
\kappa^{-1} = {\kappa_\mathrm{rad}^{-1} + \kappa_\mathrm{cond}^{-1}},
\end{align}
\citep{kippenhahn_stellar_2013}. Here, $\kappa_\mathrm{rad}$ is the radiative opacity based on the grain opacities of \citet{bell_using_1994} and the molecular opacities of \citet{freedman_line_2008}. The conductive opacity $\kappa_\mathrm{cond}$ is taken from \citet{cassisi_updated_2007}. For $\kappa_\mathrm{rad}$ we multiply the contribution of the interstellar medium grain opacity by a reduction factor of $f_\mathrm{opa} = 0.003$ following \citet{mordasini_grain-I_2014} \citep[see also][]{ormel_atmospheric_2014}. Small grains are introduced into the envelope by accreted gas, but their opacity is lessened by their coagulation and settling in planetary atmospheres. 
The opacity influences the radiative temperature gradient $\nabla_\mathrm{rad}$ which is given by 
\begin{align}
\label{eq:radgrad}
    \nabla_\mathrm{rad}=\frac{3}{64 \pi \sigma G} \frac{P}{T^4} \frac{\kappa L}{m},  
\end{align}
where $\sigma$ denotes the Stefan--Boltzmann constant, $G$ the gravitational constant, $P$ the pressure, $T$ the temperature, $L$ the luminosity, and $m$ the mass inside the local radius $r$. The radiative gradient is only applied in radiative regions, which is determined according to the Schwarzschild stability criterion as 
\begin{align}
\label{eq:SchwarzschildCriterion}
    \nabla(T, P) = \min \left(\nabla_{\mathrm{ad}}, \nabla_{\mathrm{rad}}\right),
\end{align}
where $\nabla_{\mathrm{ad}}$ is the adiabatic gradient 
\begin{equation}
    \label{eq:adiabatic_gradient_for_convection}
    \nabla_{\mathrm{ad}} = \left. \frac{\mathrm{d} \ln T}{\mathrm{d} \ln P} \right\vert_{S}.
\end{equation}
Here, $S$ is the entropy which is used in convective regions $(\nabla_{\mathrm{ad}} <
\nabla_{\mathrm{rad}})$ and is obtained directly from the tabulated equation of state of \citet{chabrier_new_2021}, which we use throughout for a constant gas mixture of $X = 0.76$ and $Y = 0.24$ with no metals ($Z=0$).

Concerning the luminosity $L$ in radiative regions (Eq.~\ref{eq:radgrad}), we use the simplification 
\begin{equation}
    \label{eq:structEq3}
    \frac{\mathrm{d} l}{\mathrm{d} r}= 0,
\end{equation}
which formally assumes a constant luminosity $L$ throughout the envelope, with $L$ originating in the core. While this is not the case in reality, the actual local value of $l$ affects the internal structure only in radiative layers, as \cref{eq:radgrad} shows, but not in convective ones. In thin outer radiative layers without significant mass, $l = L$ is a very good approximation (see Section~3.4 of \citealt{mordasini_characterization-I_2012}).
The emergence of inner radiative layers containing significant mass as found in \citet{berardo_hot_2017} does not occur for the parameters relevant to our simulations. High gas accretion rates ($\dot{M}_\mathrm{g}>\SI{E-3}{\Mearth\,\year^{-1}}$) with high shock temperatures ($T_\mathrm{sh}>\SI{2500}{\kelvin}$) are needed to reach the predicted heating regime from \citet{berardo_evolution_2017} in order to generate deep radiative layers. This is not a regime which is seen in our simulations.

For the total luminosity budget $L=\Ltot$ we follow total energy conservation, very similar as explained in \citet{mordasini_characterization-I_2012}. 
This approach inherently accounts for the core luminosity resulting from solid accretion dominant during the early formation phase and the energy release due to envelope cooling and contraction at constant mass, which is the primary source of luminosity during the evolutionary phase for giant planets. The approach leads to long-term cooling tracks during the evolutionary phase at constant mass that are in excellent agreement with conventional methods (see e.g.\ \citealt{marleau_exploring_2019}). 
We include in \Ltot also the luminosity generated by radiogenic decay $\left( L_\mathrm{radio} \right)$, which is very small \citep{mordasini_characterization-I_2012}, and from deuterium burning $\left( L_\mathrm{D\,burn} \right)$ \citep{molliere_deuterium_2012}, and we consider a potential heating contribution generated at the accretion shock by the accreting gas (details in \cref{sec:shock_conditions}). Therefore, the total luminosity is:
\begin{align}
\label{eq:Ltot}
    \Ltot &= C(L_\mathrm{M} + L_\mathrm{R}) + L_\mathrm{radio} + L_\mathrm{D\,burn} + (k-1) \Laccmax \\
    &= L_\mathrm{int} + (k-1) \Laccmax \nonumber
   .
\end{align}
Here, $C(L_\mathrm{M} + L_\mathrm{R})$ is the sum of the luminosity generated from accreting mass, i.e.\ solids and gas ($L_\mathrm{M} = L_\mathrm{M,s} + L_\mathrm{M,g}$), and from the changing radius $R$, multiplied with the correction factor $C$. This correction accounts for the negligence of the factor $\xi$ which represents the distribution of mass within the planet and its internal energy content (see \citealt{mordasini_characterization-I_2012}). 
Further, $L_\mathrm{int}$ includes luminosity due to radiogenic decay ($L_\mathrm{radio}$) and deuterium burning ($L_\mathrm{D\,burn}$). 

The approach of total energy conservation as in \citet{mordasini_characterization-I_2012} already includes the gas accretion luminosity $\Laccmax$, i.e.\ the maximal shock luminosity associated with detached gas accretion, in \Ltot and thus delivering ``hot starts'' by default as in \citet{emsenhuber_ngpps1_2021}. Thus, we introduce the shock-heating efficiency factor $k$, so that the limiting cases are the classical cold and hot starts,
\begin{align}
    L_{\text{int}} =
    \left\{
      \begin{array}{ll l}
        L-L_{\text{acc}} & \text{``Cold start''} & (k=0) \\
        L                & \text{``Hot start''}  & (k=1).
      \end{array}
    \right.
\end{align}
The $k$ factor and the cold/hot start paradigm is discussed in detail later in \cref{sec:shock_conditions}.

For the analysis of the results we also use $L_\mathrm{M,s}$, the luminosity due to the accretion of solids. To calculate it, we approximate that solids start with zero initial velocity infinitely far from the planet and reach its core. This results in a luminosity
\begin{align}
    \label{eq:lumi_solids_acc}
        L_\mathrm{M,s} = \frac{G M_\mathrm{c} \dot{M}_\mathrm{c}}{R_\mathrm{c}},
\end{align}
with $M_\mathrm{c}$ being the core mass, $R_\mathrm{c}$ the core radius and $\dot{M}_\mathrm{c}$ the solids accretion rate of the planet. We do not actually use this to calculate the total
luminosity (because it is already included in Eq.~\ref{eq:Ltot}), but it helps to break down the contributions of the total luminosity later. While in reality accreted solids are expected to dissolve in the gaseous envelope once it becomes sufficiently massive (e.g.\ \citealp{podolak_interactions_1988, helled_fuzzy_2024}), this effect is not critical for our study, since by that time the luminosity budget is dominated by gas accretion and the contribution from solid accretion has become negligible.

To analyse and compare the simulations, we also track the maximal bolometric luminosity, which is the sum of the intrinsic luminosity and the remaining luminosity from the accretion:
\begin{align}
    \label{eq:lumi_bolo}
    \Lbol = \Ltot + (1-k)  \Laccmax.
\end{align}
This corresponds to the luminosity an observer would measure in the absence of extinction or absorption.

\subsection{Boundary conditions}
\label{sec:boundaryConditions}
In the context of giant planet formation in the core accretion scenario, planets undergo three distinct phases \citep{bodenheimer_models_2000,mordasini_characterization-I_2012}. The initial formation of the still very low mass protoplanets starts in the ``attached'' phase, before transitioning into the ``detached'' phase when they accrete most of their mass. The formation process then ends in the third and last phase, the evolutionary phase at constant mass. In each of these distinct phases, a unique set of boundary conditions must be specified to solve the structure equations.

\subsubsection{The attached phase}
\label{sec:attachedPhase}
In the attached phase of planet formation, the very young planetary core with relatively low mass is still embedded in the protoplanetary disk. The planet's envelope is attached to the surrounding nebula, and from an observer's perspective, it would be impossible to see the exact boundary between the two \citep{ormel_hydrodynamicsII_2015}. Thus, the boundary condition for the pressure is given by the nebula itself with $P_\mathrm{neb}$, which we consider to be the disk's midplane pressure from the disk model \citep{pollack_formation_1996}.
The surface temperature \Tsurf of the planet is controlled by both the intrinsic temperature 
\begin{align}
    \label{eq:Tint4}
    T_\mathrm{int}^4 &= \frac{3 \tau L_\mathrm{tot}}{8 \pi \sigma \Rout^2}\\
    \tau &= \max\left(\rho_\mathrm{neb} \kappa_\mathrm{neb} \Rout, \,  2/3 \right)
\end{align}
and the disk midplane temperature $T_\mathrm{d}$ given by the time-dependent disk model. The surface temperature is obtained from the sum of the energy fluxes:
\begin{align}
    \label{eq:tempAttachedPhase}
    \Tsurf^4 = T_\mathrm{int}^4 + T_\mathrm{d}^4.
\end{align}
For the outer radius of the planet, one has to consider which part of the surrounding gas is bound to the planet. The relevant radii are the Bondi (or accretion) radius \begin{align}
    \label{eq:Bondi-radius}
    \RBondi = \frac{GM}{c_\mathrm{s}^2},
\end{align}
where $c_\mathrm{s}$ is the speed of sound in the surrounding medium, i.e.\ the disk, and the Hill radius 
\begin{equation}
    \label{eq:Hill-radius}
    R_\mathrm{H} = a \left ( \frac{M}{3 M_\star} \right )^{1/3},
\end{equation}
with $a$ being the planet's semi-major axis and $M_\star$ the stellar mass.
As in \cite{lissauer_models_2009}, the outer radius \Rout is
\begin{align}
    \label{eq:R_out}
    \Rout = \frac{\RBondi }{1+\RBondi / (\frac 14 R_\mathrm{H})},
\end{align} 
which reduces approximately to the smaller of \RBondi and $R_{\rm H}/4$ and is a first order approximation to take into account that gas in the outer parts of the Hill sphere is not bound to the planet but participates in the surrounding flow in the gas disk, as shown by hydrodynamic simulations (e.g.\ \citealp{ormel_hydrodynamicsII_2015,cimerman_hydrodynamics_2017,ali-dib_imprint_2020,moldenhauer_recycling_2022}). While there are more sophisticated approaches to deal with this effect \citep{ali-dib_imprint_2020,bailey_growing_2024,savignac_advective_2024}, we here use for simplicity to the approximation of \citet{lissauer_models_2009} because the principle that the cooling of the deeper quasi-1D part of the envelope ultimately regulates gas accretion seems robust \citep{bailey_growing_2024}.

With now a set of five given boundary conditions for $T$, $P$, \Rout, $R_\mathrm{c}$ and \Ltot, we are able to solve the initial structure equations for only one possible total planetary mass \Mpla, which consequently provides an indirect measure of the gas accretion rate. We do this by iterating on \Mpla until the correct known inner boundary condition for $M_\mathrm{c}$ is found. Details are given in \citet{mordasini_characterization-I_2012}. 

During the attached phase, the rate at which gas is accreted scales with the envelope's capacity to radiate away energy and contract (also known as its Kelvin--Helmholtz timescale $\tau_\mathrm{KH}$), allowing additional gas to flow in. Thus, the accretion rate is (not yet) limited by the disk but by the planet properties. While several parametrisations for $\tau_\mathrm{KH}$ can be found in the literature (e.g.\ \citealp{ikoma_formation_2000,mordasini_grain-I_2014}), we find that solving the structure equation directly is important, as it establishes a self-consistent link between solid and gas accretion \citep{pollack_formation_1996,kessler_interplay_2023}.

\subsubsection{The detached phase}
\label{sec:detachedPhase}
As the core mass increases during the attached phase, the rate of gas accretion gradually begins to accelerate. Eventually, the gas accretion rate surpasses the maximum supply capacity of the surrounding disk. This is when the detached phase (also called the transition stage; \citealt{bodenheimer_models_2000}) begins. In this phase, the accretion rates are known because the gas accretion rate is now limited by the disk. However, contrary to the attached phase where we iterate on the mass to find the gas accretion, we now iterate on the radius $R$ until we converge on the known total mass of the planet $\Mpla$.
We adopt the maximum disk-limited gas accretion rate \Mdotgasmax from \citet{mordasini_characterization-I_2012}, which is based on a Bondi/Hill-limited accretion regime. 

Since the planet's outermost envelope layer is no longer attached to the background nebula, the gas free-falls onto the planet. This results in additional pressure because of the accretion shock, namely the ram pressure
\begin{align}
    \label{eq:P_ram}
    P_\mathrm{ram} = \frac{\dot{M}_\mathrm{gas} v_\mathrm{ff}}{4 \pi \Rpla^2},
\end{align}
where we use $\dot{M}_\mathrm{gas} = \Mdotgasmax$ and $v_\mathrm{ff}$ is the free-fall velocity 
\begin{align}
    \label{eq:v_freefall}
    v_\mathrm{ff} = \sqrt{2G \Mpla \left( \frac{1}{\Rpla} - \frac{1}{R_\mathrm{out}} \right )}.
\end{align}
The new boundary condition for the pressure is now given by \begin{align}
    \label{eq:Pressure_total_sum_detached}
    P = P_\mathrm{neb} + P_\mathrm{ram} + P_\mathrm{Edd} + P_\mathrm{rad},
\end{align}
where we also accounted for photospheric pressure for material above the $\tau = 2/3$ layer with the Eddington expression $P_\mathrm{Edd} = {2g}/{3\kappa}$, 
and the (negligible) radiation pressure $P_\mathrm{rad} = {2 \sigma T^4}/{3c}$ where $c$ is the speed of light.
The surface temperature is again given by \cref{eq:tempAttachedPhase}. However, since $T_\mathrm{d} \lesssim \SI{100}{\kelvin}$, $T_\mathrm{int}$ rapidly becomes dominant in the detached phase, leading to
\begin{align}
        \label{eq:tempDetachedPhase}
    \Tsurf^4 = T_\mathrm{int}^4 + T_\mathrm{d}^4 \approx T_\mathrm{int}^4.
\end{align} 
Here, $T_\mathrm{int}$ is calculated again from \cref{eq:Tint4} and thus also includes the contribution from the accretion shock, using our way of calculating $\Ltot$ in \cref{eq:Ltot}.

\subsubsection{The evolution phase}
\label{sec:evolutionPhase}
The final phase begins once the gas disk around the planet vanishes, letting it evolve at a constant mass. The discontinuation of gas accretion also means that there is no resupply of grains \citep{mordasini_grain-I_2014}. Consequently, we gradually reduce $f_\mathrm{opa}$ to 0 once $\dot{M}_\mathrm{gas} \leq \SI{E-5}{\Mearth \, \year^{-1}}$. \citet{mordasini_grain-I_2014} formally predict no dependence of grain opacities on the gas accretion rate. However, within that model, it is clear that when gas accretion (virtually) stops and thus the source of grain supply stops, $f_\mathrm{opa}=0$. To achieve a smooth transition of the $f_\mathrm{opa}$ factor, we start to reduce it below the formation-phase value until $f_\mathrm{opa}=0$ for $\dot{M}_\mathrm{gas} \leq \SI{E-7}{\Mearth \, \year^{-1}}$. The chosen limits correspond to frequently occurring lower limits of the gas accretion rate reached shortly before disk dispersal in our simulations.

Furthermore, for $L_\mathrm{acc} = 0$ we have $\Ltot = \Lint$ resulting in 
\begin{equation}
\label{eq:StefanBoltzmannLaw}
    T^4_\mathrm{int} = \frac{\Lint}{4 \pi \sigma \Rpla^2}.
\end{equation}
After the dispersal of the disk, the planet is exposed to stellar irradiation. A simple approximation for weakly irradiated objects for the surface temperature is then given by 
\begin{equation}
    \label{eq:Tsurf-evolution}
        \Tsurf^4 = T^4_\mathrm{int} + (1-A)T^4_\mathrm{eq},
\end{equation}
where $A = 0.343$ is the geometric albedo, fixed to that of present-day Jupiter \citep{guillot_interiors_2005}, and 
\begin{equation}
    T_\mathrm{eq} = T_\star \sqrt{\frac{R_\star}{2a}}
\end{equation}
is the equilibrium temperature. The stellar effective temperature $T_\star$ and radius $R_\star$ evolve according to the tracks from \citet{baraffe_new_2015}.
In the evolution phase, the pressure boundary condition in \cref{eq:Pressure_total_sum_detached} reduces to 
\begin{equation}
    P = P_\mathrm{Edd} + P_\mathrm{rad}.
\end{equation}

\subsubsection{Shock conditions -- hot or cold start?}
\label{sec:shock_conditions}

Two decades ago, \citet{fortney_youngJupiters_2005} and \citet{marley_luminosity_2007} pointed out that the energy transfer at the accretion shock in the detached phase plays a key role in setting the ``initial'' entropy or luminosity of planets, that is, those values at the beginning of the post-formation cooling. If the energy of the accreting gas is fully absorbed into the planet, a so-called ``hot start'' ensues, and conversely, if the energy is fully radiated away at the accretion shock, the scenario is referred to as a ``cold start''. 
These limiting cases can lead to strong differences in the predicted luminosity or brightness of young gas giants, especially for more massive ones, with cold starts leading to much fainter planets post-formation \citep{marley_luminosity_2007,fortney_synthetic_2008}. This has major consequences when interpreting observations (e.g.\ \citealp{spiegelBurrows_spectral_2012,marleau_constraining_2014}).

Reality can lie anywhere in the spectrum between the two extremes, called a ``warm start''.
In fact, this ambiguity has been extensively addressed observationally, and the coldest starts are incompatible with direct-imaging results (e.g.\ \citealp{nielsen_gemini_2019,vigan_SPHERE-III_2021,dupuy_limits_2022}).
When modelling gas accretion and the internal structure in the detached phase (e.g.\ \citealp{mordasini_characterization-I_2012,mordasini_luminosity_2013,berardo_evolution_2017,berardo_hot_2017,cumming_primordial_2018,emsenhuber_ngpps1_2021}), the shock efficiency is commonly expressed by the parameter $\eta\in[0,1]$, with $\eta = 1$ referring to cold start and $\eta = 0$ to hot start scenarios (see \citealp{marleau_accretionShock-I_2017} and references therein).
A similar question exists in the context of star formation (e.g.\ \citealp{prialnik_outcome_1985,hosokawa_evolution_2009,commerCon_physical_2011,baraffe_observed_2012,vaytet_simulations_2013,geroux_multi_2016,baraffe_selfconsistent_2017,bhandare_turn_2025}).

In this work, for the first time, we do not assume a constant value of the shock efficiency $\eta$ but rather calculate it dynamically as a function of the evolving planet and system properties. This is motivated by the fact that $\eta$ is not directly constrained for individual systems and is expected to depend on the accretion flow and shock conditions, which evolve during runaway gas accretion. We couple our interior structure code to semi-analytical expressions of the shock heating efficiency, which are based on and backed by detailed radiation-hydrodynamic simulations resolving the energy transfer at the accretion shock.
\citet{mordasini_characterization-III_2017} showed that, when the ``core mass effect'' \citep{mordasini_luminosity_2013} is allowed to operate, adopting a constant extreme value $\eta=0$ or $\eta=1$ does not necessarily translate into large differences in the post-formation luminosity, and that reproducing the coldest starts of \citet{marley_luminosity_2007} requires rather extreme assumptions.
Here, however, we purposely choose formation pathways with relatively modest core masses (see below) to minimise core-mass-driven self-heating, so that differences caused by the shock treatment are not automatically washed out. Moreover, a varying shock efficiency is physically motivated because the shock conditions (in particular $\dot{M}_\mathrm{gas}$, $\Rpla$, and the pre-shock Mach number) evolve rapidly during runaway accretion; even if the final post-formation luminosity converges, the preceding \LT track can differ, which is what we aim to capture.
The one-dimensional, spherically-symmetric simulations of \citet{marleau_accretionShock-I_2017,marleau_accretionShock-II-2019} show that, at the accretion shock, only a small fraction of the total available kinetic energy of the accreting gas actually passes the shock and ends up directly heating the planet. However, especially with high gas accretion rates, the outgoing radiation at the shock can be a considerably large energy flux, even surpassing $\Lint$ (see Sect.~\ref{sec:compColdHotStart}). This radiation preheats the infalling gas, increasing the energy ultimately advected into the planet to non-negligible levels.
It is easy to estimate that already a minor fraction of the total shock luminosity heating the planet can represent a significant energy flux compared to extreme cold accretion (see Fig.~11 of \citealp{marleau_accretionShock-II-2019}).
This is the case also observationally, where differences between ``hot'' and ``very hot'' are not significant because of the short associated KH timescale after which the two scenario become the same. In contrast, the difference between a warm and extreme cold start can remain for several million years \citep{marley_luminosity_2007,spiegelBurrows_spectral_2012}, affecting the statistics of planetary luminosities \citep{mordasini_characterization-III_2017} and the interpretation of direct-imaging discoveries (e.g.\ \citealp{samland_spectral_2017,marleau_exploring_2019,vigan_SPHERE-III_2021}).

Following \cite{marleau_accretionShock-II-2019}, the amount of accretion energy that results in heating up the planet and therefore contributes to its total luminosity is 
\begin{equation}
\label{eq:Qshock+}
\Qshock = \left(\frac{2}{(\Gamma_1-1) \Mach^2}+\frac{1}{\Gamma_1^2 \Mach^4}\right) \Laccmax,
\end{equation}
where \Laccmax is the full shock luminosity corresponding to the kinetic energy of the accreted free-falling gas:
\begin{equation}
\label{eq:Laccmax}
 \Laccmax = G \Mpla \dot M_\mathrm{gas} \left( \frac{1}{\Rpla} - \frac{1}{R_\mathrm{out}} \right).
\end{equation}
This expression for $Q^+_\mathrm{shock}$ is valid in the ``isothermal'' limit, in which the equilibrium gas temperature is the same up- and downstream of the shock. This was found always to hold for the planet-formation shock \citep{marleau_accretionShock-I_2017,marleau_accretionShock-II-2019}, which is supercritical (e.g., \citealt{drake_high_2006,commerCon_physical_2011}).
We calculate the first adiabatic index $\Gamma_1$ and the mean molecular weight $\mu$ for the outermost envelope layer (which is just downstream of the accretion shock) with the EOS.
Then, $\Mach = v / c_\mathrm{s}$ is the Mach number of the accretion flow just before the shock, with $v = v_\mathrm{ff}$ from Equation~(\ref{eq:v_freefall}) and the speed of sound
\begin{equation}
    \label{eq:soundspeed}
    c_\mathrm{s} = \sqrt{\Gamma_1\frac{k_\mathrm{B} \Tsh }{\mu \amu}},
\end{equation}
with \kB the Boltzmann constant and \amu the atomic mass unit.

In \cref{eq:v_freefall} we take into account that the gas falls from \Rout onto the planet\footnote{Because the planet is collapsing quickly but still quasi-statically \citep{mordasini_characterization-I_2012}, $\Rout \gg R$ most of the time. However, shortly after detachment, the ${\Rout^{-1}}$ term can be significant for a short time.}, and not from infinity as is assumed for the planetesimals.
The temperature at the boundary condition for the shock is formally given by an implicit equation \citep[Eq.~32a in][]{marleau_accretionShock-II-2019}:
\begin{equation}
\label{eq:implicit_Tshock}
\Tsh^4=\frac{L_{\text {dnstr}}+\eta^{\text{kin}}\left( \Tsh \right) \Laccmax }{16 \pi \Rpla^2\sigma \fred^+ \left( \Tsh \right)},
\end{equation}
where $\eta^\textnormal{kin}$ is the kinetic efficiency at the shock \citep{marleau_accretionShock-I_2017} and $\fred^+$
is the reduced flux upstream of (``above'') the shock. The reduced flux, also called the ``streaming factor'', indicates the degree to which the radiation is freely streaming ($\fred \rightarrow 1$) or diffusing $(\fred\rightarrow0$). We assume that the shock is always in the free-streaming limit with $\kappa \rho \Rpla \ll 1$.
This is a good approximation throughout most of the detached phase, since we find to have high Mach numbers $\Mach$ and a low $k$ with our approximation of having spherically symmetric gas accretion from \Rout. Thus, Equation~(\ref{eq:implicit_Tshock}) becomes
\begin{equation}
   T_\mathrm{sh}^4 = \ell \left ( \frac{G}{16\pi \sigma} \right ) \frac{\Mpla \dot{M}}{\Rpla^{3} },
    \label{eq:Tshock4}
\end{equation}
where $\ell = 1 + L_\mathrm{dnstr} / \Laccmax$, and $L_\mathrm{dnstr}$ is the sum of all luminosities originating from inside the planet:
\begin{align}
L_\mathrm{dnstr} &=  \Ltot\\
&\approx \Lint - \Laccmax  \qquad (\textnormal{with }k\approx0)
\end{align}

\begin{figure}
\centering
\includegraphics[width=0.9\hsize]{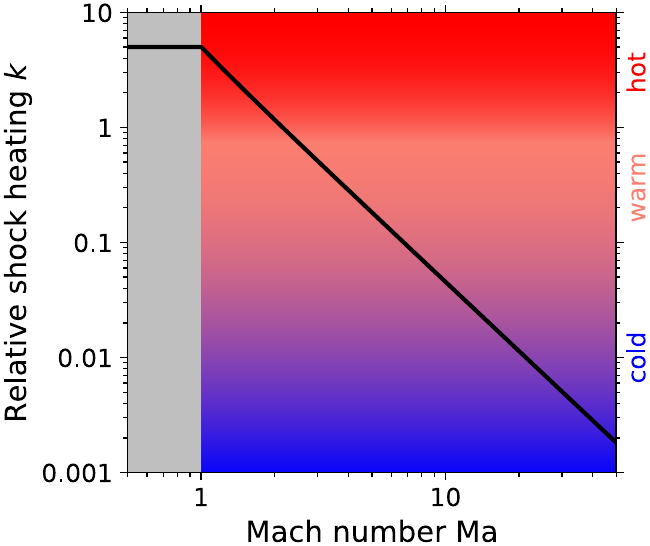}
   \caption{
   Relative shock heating $k$ (Eq.~\ref{eq:kFactor}) as a function of Mach number $\Mach$ for a constant $\Gamma_1 = 1.44$. For $\Mach < 1$ there is no accretion shock, thus we artificially set $k(\Mach < 1)=k(\Mach=1) =5.16$. For $\Mach > 1$ we identify three regions: 
   For $k \sim 0$ the gas accretion resembles a cold start scenario, while $k = 1$ represents a classical hot start scenario. Everything in between is labelled as a warm scenario, although the transition around $k\approx 0.1$ is chosen arbitrary.
   }
      \label{fig:kfactor}
\end{figure}

We now define the $k$-factor as the relative shock heating, representing the amount of gas accretion energy that ends up in the planet, scaled to the maximum gas accretion luminosity \Laccmax:
\begin{align}
 k = 
 \begin{cases}
     \displaystyle \frac{\qshock}{\Laccmax} = \frac{2}{(\Gamma_1 - 1) \Mach^2 } + \frac{1}{\Gamma_1^2 \Mach^4},& \Mach > 1 \\
     \displaystyle k_\mathrm{max} = \frac{2}{\Gamma_1-1} + \frac{1}{\Gamma_1^2},
     & \Mach \leq 1.
 \end{cases}
 \label{eq:kFactor}
\end{align}
Fig.~\ref{fig:kfactor} shows $k(\Mach)$ for a value of $\Gamma_1 = 1.44$, and it can be seen that for $\Mach \rightarrow 1$ the classical hot-start limit of $\eta=0$, corresponding to a relative shock heating of $k=1$, does not hold anymore. At first it might be surprising that more energy can be accreted into the planet than that available from the maximum accretion luminosity. However, this can be explained with the fact that the $\qshock$ expression from \cref{eq:Qshock+} also includes the enthalpy of the gas before the accretion itself and not only the kinetic energy it brings from the accretion process\footnote{To see why the enhancement can be as large as a factor of ${\sim}5$, consider the limiting case where warm gas is accreted from just outside the planet's surface, $\Rout \approx \Rpla$. In this limit, $\Laccmax \rightarrow 0$, yet the accreted gas still carries significant thermal energy in the form of enthalpy. Since $k = \qshock/\Laccmax$, the ratio diverges as $\Rout \rightarrow \Rpla$, showing that $k \gg 1$ is not only possible but expected whenever the gas enthalpy dominates over the kinetic energy gained during infall. A factor of six is therefore not surprising; it simply reflects that for the physical conditions considered here, the enthalpy contribution exceeds the accretion luminosity but not vastly.} \citep{marleau_accretionShock-II-2019}.
For $\Mach \leq 1$ there is no accretion shock and it is unclear how to treat the heating from accretion. For historical reasons we keep the heating at a (for a given $\Gamma_1$ value) fixed $k_\mathrm{max}$ value (see Eq.~\ref{eq:kFactor}).  
Ultimately, the $k$-factor acts like a variable $\eta$-factor, enabling us to compute the total luminosity (and thus entropy) of the planet as stated in \cref{eq:Ltot}. 

\begin{figure}
\centering
\includegraphics[width=0.9\hsize]{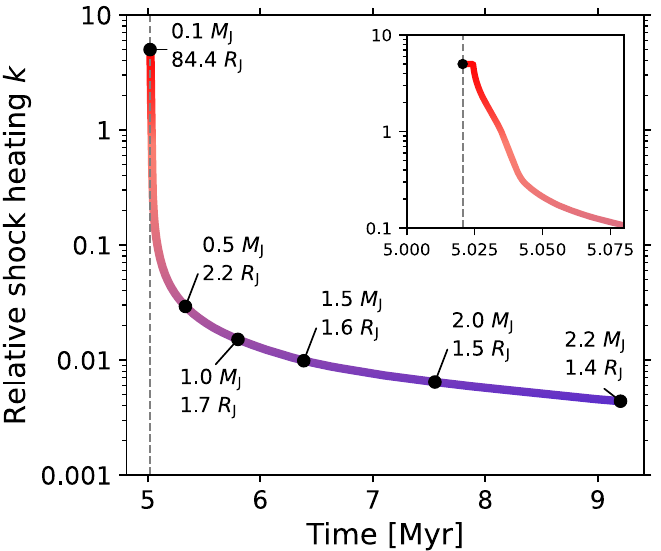}
   \caption{Evolution of the relative shock heating $k$ as a function of time during the gas accretion of the forming giant planet in the detached phase. The data shown is from the default case planet from \cref{sec:defaultCase}. The mass and radius evolution can be seen by their values at selected data points. The inset highlights the brief hot phase after detachment.}
      \label{fig:kFactorOverTime}
\end{figure}

In this paper, we intentionally keep the core masses relatively low with $M_\mathrm{c} \lesssim 25\, \Mearth$ to avoid large core masses, which by the ``core mass effect'' lead automatically to hot starts \citep{mordasini_luminosity_2013}. We relax this in \cref{sec:migratingPlanets}. Obtaining colder starts makes it possible to compare with \citet{marley_luminosity_2007}. 
Initial conditions are therefore chosen to avoid the hot-start effect induced by massive solid cores, so that the cold-/hot-start ambiguity can be examined without interference from core-related influence.

Our approach couples for the first time the results of a model for the radiative efficiency of the planetary gas accretion shock with a full planet formation model calculating the planets' interior structure at each moment during the formation phase, making it possible to obtain the post-formation thermodynamic state. However, it is clear that this work still has important limitations: as mentioned in the introduction, we focus in this paper on the planet itself and how the basic mechanisms of planetary formation express themselves in the HR diagram. Thus, we use in this work the spherically-symmetric approximation and neglect the presence of a CPD and associated more complex gas accretion geometries like magnetospheric \citep{hartmann_magnetosphericAccretion_1994, muzerolle_magnetosphericAccretion_1998, thanathibodee_magnetosphericAccretion_2019} or boundary layer accretion \citep{lynden-bell_evolutionViscousDiscs_1974, ghosh_accretion-II_1979}. Similarly, we leave it up to future work to study how the forming planets would appear observationally via direct imaging and other disk observations (e.g., \citealt{szulagyi_observability_2019,chen_observability_2022,choksi_spectral_2025,taylor_2D_2026}).

\begin{table*}[th]
    \caption{Initial conditions and settings for the default in-situ simulation case which results in a planet with $M_\mathrm{tot} = 2.2 \, M_\mathrm{J}$. }
    \label{tab:modelSetup}
    \centering
\begin{tabular}{clcc}
\hline \hline
Quantity   & Description or effect                       & Default value                 & Variation \\
\hline %
$a$                                         & Semi-major axis                   & 5.2 \si{\astronomicalunit}                           &       \cref{sec:migratingPlanets}    \\
$\Sigma_{\mathrm{s}, 0}$           & Initial planetesimal surface density at $\SI{5.2}{\astronomicalunit}$ & 10 \si{\g \per \square \cm}&       \cref{sec:twoPopComparison}       \\
$M_{\mathrm{disk},0}$                                   & Initial total disk (gas) mass     & 0.04~$\Msun$                        &       \cref{sec:higherMassesComparison,sec:twoPopComparison} \\ 
$\alpha$                                                                    & \citet{shakura_black_1973} disk viscosity parameter          &    $10^{-3}$                  &       {--}     \\
$\beta_\mathrm{gas}$                                                        & Initial gas disk power law index  &    0.9                        &       {--}     \\
$\beta_\mathrm{pla}$                                                        & Planetesimal disk power law index &    1.5                        &       {--}     \\
{$f_\mathrm{opa}$} & Grain opacity reduction factor & 0.003 (0 during evolution)        & \cref{sec:hayashiTracks} \\
$k$ & Relative shock heating (Eq.~\ref{eq:kFactor}) &  variable \citep{marleau_accretionShock-II-2019}  &      \cref{sec:compColdHotStart}      \\
$\dot{M}_\mathrm{wind} $             & External photo-evaporation rate   &   $10^{-7}$ \si{\Msol\,\year^{-1}} &    \cref{sec:twoPopComparison}    \\ 
{$(X,Y,Z)$} & Gas composition (H, He, metal mass fractions)                                                               & $X=0.76$, $Y=0.24$, $Z=0$            &     {--}    \\ 
{--} & Initial embryo mass $            $                                            & 0.01 $\Mearth$                              &     {--}      \\
{--} & Migration                                                                               & not included                      &     \cref{sec:migratingPlanets}        \\
{--} & Disk evolution                                                                          & included                          &     {--}      \\
{--} & Planetesimal ejection                                                                  & included                          &     {--}      \\
{--} & Planetesimal size                                                                      & 10~km                             &     \cref{sec:twoPopComparison,sec:migratingPlanets}       \\
{--} & Fate of dissolved planetesimals                                                        & sink to core interface            &     {--}      \\
{--} & Atmospheric boundary condition                                                                             & Eddington approximation                &      {--}     \\      \hline
\end{tabular}    
\end{table*}

\section{Imprint of formation mechanisms in planetary HR diagrams}
\label{sec:Methods+main+var}

In this section, we outline our simulation setup and present the resulting planetary HR diagrams in order to study how the mechanisms of planetary formation manifest themselves in HRDs. We begin by explaining the default setup (Sect.~\ref{sec:defaultCase}) and study in depth a single formation simulation leading to a ${\sim}2\,\Mjup$ planet as our default case (\cref{tab:defHRD-data} and \cref{fig:defHRD}), examining the principal characteristics of the \LT track throughout the formation and evolution of the planet over a period of \SI{10}{\giga \years} (Sect.~\ref{sec:resultsdefaultcase}). In the subsequent sections, we explore the influence of various parameters and assumptions for the formation process and their resulting HRD such as $\eta$ for purely hot and cold starts (\cref{sec:compColdHotStart}) or the influence of contraction (\cref{sec:hayashiTracks}) in the detached phase, different planetary masses (\cref{sec:higherMassesComparison}), a core accretion scenario with pebbles instead of planetesimals (\cref{sec:twoPopComparison}) and lastly the influence of orbital migration (\cref{sec:migratingPlanets}), all for simulations over a period of \SI{200}{\mega \years}.

\subsection{Simulation setup for the default case}
\label{sec:defaultCase}

\begin{table}[!ht]
\caption{Planet mass and radius at selected times, marked in \cref{fig:defHRD}.}
\label{tab:defHRD-data}
\centering
\begin{tabular}{cllll}
\hline
\hline
Symbol & Event          & Time & Mass                   & Radius              \\ \hline
\hspace{-1.5mm} $\pmb{\bullet}$                                           & ``Start''        & $\SI{0.08}{\mega \years}$ & $2 \, M_{\mathleftmoon}$     & ${0.4}\,{\Rjup}$  \\
\hspace{-1.45mm} $\bigstar$                                                & Detachment     & $\SI{5.0}{\mega \years}$  & ${0.1}\,{\Mjup}$             & ${84.4}\,{\Rjup}$ \\
\hspace{-1.45mm} \scalebox{1.5}[1]{\raisebox{0.0pt}{$\blacktriangledown$}} & Maximum $L$     & $\SI{6.1}{\mega \years}$  & ${1.3}\,{\Mjup}$             & ${1.7}\,{\Rjup}$  \\
\hspace{-1.2mm} $\blacksquare$ & Disk dispersal & $\SI{9.2}{\mega \years}$  & ${2.2}\,{\Mjup}$ & ${1.4}\,{\Rjup}$  \\
\hspace{-1.2mm} \textbf{+}                                                 & Jupiter age    & $\SI{4.6}{\giga \years}$  & ${2.2}\,{\Mjup}$             & ${1.1}\,{\Rjup}$\\
\hline
\end{tabular}
\end{table}

In \cref{tab:modelSetup}, we show the initial conditions and settings for our default simulation. Following the example of past works on giant planet formation \citep{pollack_formation_1996, lissauer_models_2009,dangelo_growth_2021}, and to avoid excessive complexity introduced by additional input physics, our default scenario is based on a single embryo, planetesimal accretion, and in situ formation. The embryo is placed at $t=\SI{0}{\years}$ at a semi-major axis of $a=\SI{5.2}{\astronomicalunit}$, where it will remain, in a gas disk with an initial total mass of $\SI{0.04}{\Msol}$ around a solar-type star. We use an independent and simplified MMSN-profile for the disk of planetesimals:
\begin{equation}
\label{eq:Sigmasimal}
    \Sigma_\mathrm{pla}(r) = \Sigma_\mathrm{s,0} \left( \frac{r}{\SI{5.2}{\astronomicalunit}}\right) ^{-\beta_\mathrm{pla}}
\end{equation}
We neglect the impact of icelines, which allows for a better comparison later in \cref{sec:higherMassesComparison}, where we will increase the initial total disk mass $M_{\mathrm{disk},0}$ to study the formation of planets of different masses. We leave the value and structure of $\Sigma_\mathrm{pla}$ unchanged in most of the coming sections in order to achieve low and comparable core masses.

\subsection{Results for the default case}
\label{sec:resultsdefaultcase}

\begin{figure*}
        \centering
        \includegraphics[]{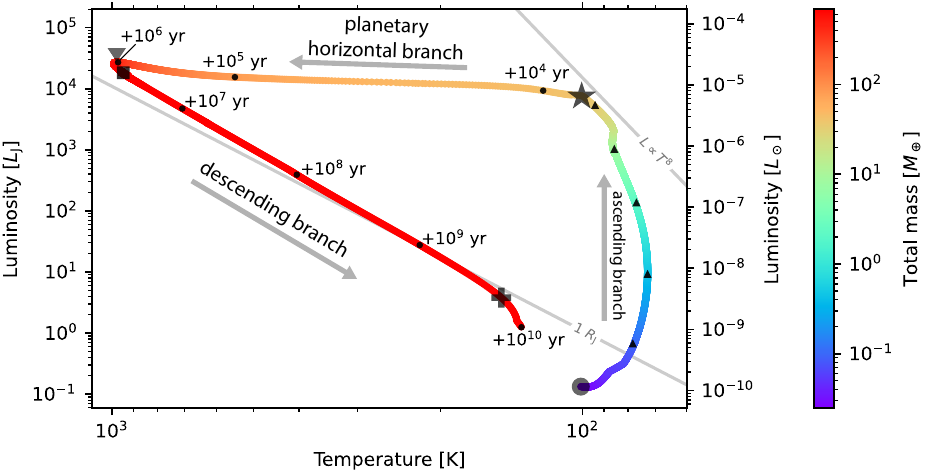}
        \caption{
    HR diagram track of the default case (final mass of 2.2~$\Mjup$; see Table~\ref{tab:modelSetup}), showing its surface temperature (Eqs. \ref{eq:tempAttachedPhase}, \ref{eq:tempDetachedPhase} or \ref{eq:Tsurf-evolution}, depending on the phase) and total luminosity (Eq. \ref{eq:Ltot}). Highlighted key moments show the start point (\scalebox{1.3}{\raisebox{-0.75pt}{$\bullet$}}), detachment ($\bigstar$), the point of maximum luminosity (\scalebox{1.3}[1]{\raisebox{2pt}{$\blacktriangledown$}}), disk dispersal ($\blacksquare$), and the age of Jupiter (\scalebox{1.2}{$\textbf{+}$}). \cref{tab:defHRD-data} lists the corresponding values for time, mass and radius. In the ascending branch small black triangles ($\blacktriangle$) show linearly spaced time indicators every Myr. The black dots ($\bullet$) represent post-detachment intervals equally spaced in log(time). Grey reference lines show two trends: the predicted slope derived in \cref{appendix:appendixB_slopeAttachedPhase}, evaluated with the median late-attached-phase surface density of $\Sigma_\mathrm{pla} = \SI{1.7}{\gram \per \square \cm}$ (labelled with $L\propto T^8$), and a constant-radius track for $\Rpla = 1\,{\Rjup}$, which approximates a cooling planet in the evolutionary phase before stellar irradiation becomes relevant (last downturn). 
    }
        \label{fig:defHRD}
\end{figure*}


\begin{figure*}
        \centering
        \includegraphics[width=0.95\linewidth]{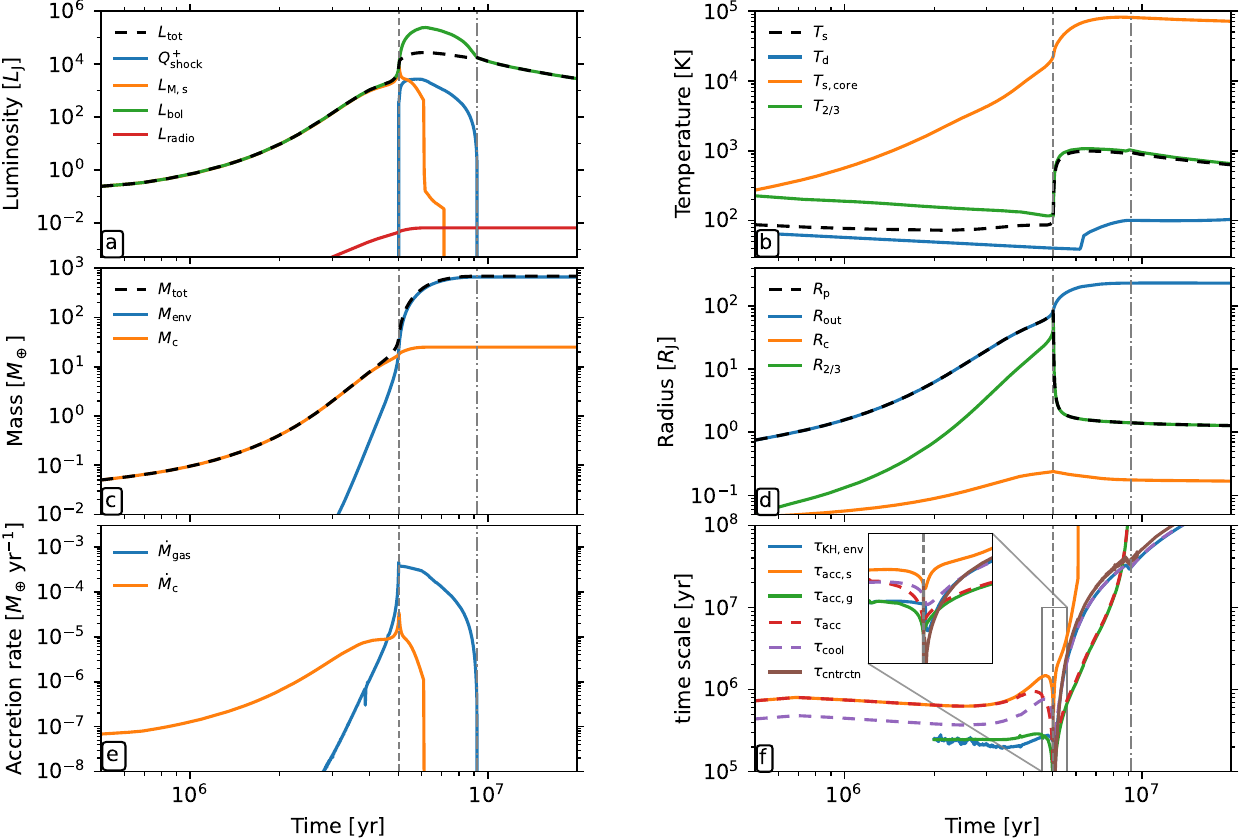}
        \caption{
Time evolution, up to \SI{20}{\mega\years}, of selected quantities related to (a)~luminosity, (b)~temperature, (c)~mass, (d)~radius, (e)~mass accretion, and (f)~time scales. In each panel, a dashed (dot-dashed) line marks the moment of detachment (disk dispersal). Details: (a)~Total luminosity emerging from the planet surface \Ltot (without the fraction of the shock luminosity that is radiated away in the accretion shock), the part of the gas accretion shock luminosity that heats the planet $\Qshock$, the solids accretion luminosity $L_{\rm M,s}$, the bolometric luminosity \Lbol and the luminosity from radioactive decay $L_{\rm radio}$. (b)~The planet surface temperature $T_{\rm s}$, the disk midplane temperature $T_{\rm d}$, the core surface temperature $T_{\rm s,core}$ and the temperature $T_{2/3}$ at Rosseland optical depth $\tau = 2/3$. (c)~The total planet mass $M_{\rm tot}$, the envelope mass $M_{\rm env}$, and the core mass $M_{\rm c}$. (d)~The planet radius $\Rpla$, the total outer radius \Rout (see Eq.~\ref{eq:R_out}), the core radius $R_{\rm c}$, and $R_{2/3}$, the radius at optical depth $\tau = 2/3$. (e)~The gas and solids accretion rates $\dot{M}_{\rm gas}$ and $\dot{M}_{\rm c}$, respectively. (f)~The Kelvin--Helmholtz time scale of the envelope $\tau_{\rm KH,\,env} = E_{\rm env,\,tot} / \dot{E}_{\rm env,\,tot}$, the solids accretion time scale $\tau_{\rm acc,\,s} = M_{\rm c}/\dot{M}_{\rm c}$, the gas accretion time scale $\tau_{\rm acc,g} = M_{\rm env}/\dot{M}_{\rm gas}$, the mass doubling time scale $\tau_{\rm acc} = \Mpla / \dot{M}_{\rm p}$, the cooling time scale $\tau_{\rm cool} = |E_{\rm tot}| / L_{\rm tot}$, and the contraction time scale $\tau_{\rm cntrctn} = \Rpla/\dot{R}_{\rm p}$.
}
    \label{fig:defaultCasePlots}
    \end{figure*}

\Cref{fig:defHRD} shows the track of the default case in the HR diagram, plotting the total luminosity of the planet \Ltot against the temperature at the planet surface $T_{s}$. 
We only show data for $\Mpla>\SI{0.025}{M_\oplus}$ ($\sim2$~lunar masses) in the HRD. Data points below this value are influenced by details of the initialisation of the simulation, such as the assumed dynamical state of the planetesimals. 
Note that \Ltot excludes the shock luminosity radiated away, so neither the temperature nor the luminosity in this diagram accounts for this fraction (see \cref{sec:compColdHotStart} for a comparison).
Except for the early attached phase, where the contribution of the nebular gas temperature is important, Eq.~(\ref{eq:tempAttachedPhase}) then simplifies to $\Ltot=4 \pi \Rpla \sigma T_{s}^4$. The accretion shock radiation is likely emerging only from a small part of the surface and is characterised by a different spectral energy distribution than that of the overall planet surface. It is, however, included and visible in \Lbol.

The temporal evolution of key quantities, such as luminosity, temperature, mass, radius, accretion rates, and timescales, is detailed in \cref{fig:defaultCasePlots}. These quantities elucidate the physical processes underlying the planet's formation and evolution track in the HRD.

A defining feature of the tracks during the formation with planetesimal accretion is the presence of three distinct branches, each corresponding to one of the phases outlined in \cref{sec:boundaryConditions}: attached, detached, or evolutionary.
Based on the direction that the planets evolve along these three branches, we call those regions the ``ascending branch'', the ``planetary horizontal branch'' and the ``descending branch''.

\subsubsection{The ascending branch}
\label{sec:ascendingBranch}

The formation process of the young planet starts in the bottom right corner of the HRD (Fig.~\ref{fig:defHRD}) with low luminosity and surface temperature. The planet is in the attached phase and is still deeply embedded in a relatively massive protoplanetary disk. It gains almost all its new mass from solid accretion (Fig.~\ref{fig:defaultCasePlots}e), so that $L \approx L_\mathrm{M,s}$. Combined with the relatively large total radius, the low initial luminosity naturally leads to a cold intrinsic temperature $T_\mathrm{int}$ which is only marginally hotter than the early midplane temperature of the disk $T_\mathrm{d}$. Thus, at least initially, the contribution of the disk to the surface temperature of the planet dominates in \cref{eq:tempAttachedPhase}. Additionally, the disk temperature also directly influences the planet's outer radius. For lower mass planets, $\Mpla \approx M_\mathrm{c}$, and $\Rpla \approx \RBondi$ \citep{lissauer_models_2009}, so that \begin{align}
    L &= 4 \pi \RBondi^2 \sigma \left(T_\mathrm{int}^4 + T_\mathrm{d}^4\right) \\
    \label{eq:lumiForReqRacc}
    &= 16 \pi \frac{G^2 M_\mathrm{c}^2}{\chi^2}\sigma \left( \frac{T_\mathrm{int}^4}{T_\mathrm{d}^2} + T_\mathrm{d}^2 \right),
\end{align}
where we defined for $c_\mathrm{s}^2$ in \RBondi (Eq.~\ref{eq:Bondi-radius})
\begin{equation}
\label{eq:cs2defchi}
c_\mathrm{s}^2 = \frac{\Gamma_1 k_\mathrm{B} T_\mathrm{d}}{\mu \amu} \equiv \chi T_\mathrm{d}. 
\end{equation}
Consequently, the initial slope of the ascending branch is still strongly influenced by the cooling disk, and the accretion of solids onto the small core is not enough to result in a considerable heating of the large and rapidly growing radius of the embedded young planet.

For even larger radii at a later time in the attached phase, the outer radius transitions from $\RBondi$ to $\frac 14 R_\mathrm{H}$ \citep{lissauer_models_2009}. Around that time, the influence of the disk starts to diminish because its temperature does not set the radius anymore and for $T_\mathrm{int} > T_\mathrm{d}$, \cref{eq:tempAttachedPhase} reduces to $T \approx T_\mathrm{int}$. By assuming a shear-dominated velocity distribution for the planetesimals at this stage where $\Rout \approx \frac 14 R_\mathrm{H}$, it is possible to derive a semi-analytical approximation of the luminosity--temperature relationship, which yields $L \propto T^8/\Sigma_\mathrm{pla}$ (see \cref{sec:appB-Part2-lateAttachedSlope}). \cref{fig:defHRD} shows this trend for $\Sigma_\mathrm{pla} = \SI{1.7}{\gram \per \square \cm}$, the median value of the surface density in the last Myr before the planet detaches. The same slope can be observed for planets of various masses in the late attached phase later in \cref{fig:HRD_higherMasses}.

In the last ${\sim}1$~Myr before detachment, the gas accretion rate $\dot{M}_\mathrm{gas}$ finally exceeds the solids accretion rate $\dot{M}_\mathrm{c}$. 
\cref{fig:defaultCasePlots}f shows that the ascending branch is dominated by $\tau_{\rm acc} \approx \tau_{\rm acc,s}$, which also dictates $\tau_{\rm cool}$. Even though $\tau_{\rm acc,g}$ is much shorter, the envelope mass is still orders of magnitudes lower than the core mass, thus letting $\tau_{\rm acc,s}$ dominate.
Only later, when $\tau_{\rm acc} \approx \tau_{\rm acc,g}$, can the envelope cool (and contract) fast enough so that $\dot{M}_{\rm gas}$ exceeds the maximum disk supply rate and the planet eventually detaches.

In the late ascending branch, $\tau_{\rm cool} < \tau_{\rm KH,env}$ because solid accretion still generates most of the luminosity and thus $\Ltot \approx L_{\rm M,s}$. 
The classical picture of $\tau_{\rm KH,env}$ defining the gas accretion rate \citep{ikoma_formation_2001} does not apply to our case due to the ongoing accretion of solids: the envelope of the planet becomes more massive primarily via the increase of the (attracting) core mass and the associated expansion of the Hill radius, rather than through the cooling process of the envelope at constant core mass which allows new gas to stream into the envelope.

\subsubsection{The planetary horizontal branch}
\label{sec:horizontalBranch}

The detachment of the planet and the onset of the planetary horizontal branch\footnote{We explicitly use the term ``planetary horizontal branch'' to avoid confusion with the stellar horizontal branch in HR diagrams, which arises from entirely different physical processes.} shows up as a sharp knee in the HRD track in \cref{fig:defHRD}. The surface of the planet is now no longer connected to the background nebula and the radius starts to contract quickly \citep{bodenheimer_models_2000}. From directly before to directly after detachment, the solid accretion rate goes through a sharp maximum because of the extension of the solid feeding zone which is in turn due to the rapidly increasing mass. Shortly after detachment, the solid accretion rate decreases strongly as scattering rather than accretion occurs and because the planetesimal capture radius becomes rapidly smaller \citep{mordasini_characterization-I_2012, podolak_detailedCalculations_2020}. This leads via a self-amplifying mechanism to the effect that massive cores lead to hotter (higher-entropy) planets \citep{mordasini_luminosity_2013}.
The contraction timescale $\tau_\mathrm{cntrctn}$ equals the cooling timescale $\tau_\mathrm{cool}$ from the point on where $L_{\Qshock} > L_\mathrm{M,s}$, about $\SI{50}{\kilo \years}$ after detachment. This means that contraction is limited by the ability to radiate away energy, that is by the cooling of the envelope.

The associated horizontal movement in this phase is extremely fast; in only \SI{0.1}{\mega\years}, the surface temperature of the planet increases from 100~to \SI{600}{\kelvin}. This contraction is what defines the planetary horizontal branch of the formation track in the HR diagram, as it is the main reason for the heating of the envelope for cold-start-like scenarios. We will see in \cref{sec:hayashiTracks} how contraction alone can define the heating and cooling of the envelope without the effect of mass accretion, and thus, be the main factor of the horizontal movement in the HRD.

In these early stages of the horizontal branch, the solid accretion and consequently also $L_\mathrm{M,s}$ decrease significantly due to the strongly decreasing planetesimal capture radius of the planet and to the associated scattering. Simultaneously, the gas accretion rate peaks and stays high for a prolonged period thereafter. The gas accretion is now limited by the disk and its gas reservoir.

From \cref{fig:kfactor} and the detailed comparison to classical hot- and cold-start scenarios in \cref{sec:compColdHotStart} it becomes apparent that our default case can be classified somewhere between a classical (or ``extreme'') cold start from \citet{marley_luminosity_2007} and a warm start.
The $k$-factor stays in the hot or warm region ($k \gtrsim 0.25$) for only ${\sim}\SI{25000}{\years}$, a time for which the free fall distance is still relatively short ($\Rout$ is not much larger than $\Rpla$) and thus Mach numbers are low. For the remaining ${\sim}\SI{4}{\mega \years}$ the $k$-factor is small and the accretion is close to, but not completely, a cold-start scenario. This result agrees with the result predicted by \cite{marleau_accretionShock-II-2019} where high gas accretion rates ${\ge} 10^{-3} \, \Mearth \mathrm{yr}^{-1}$ are needed for hot starts. The method we use keeps the shock heating relatively low due to the modest gas accretion rate and the high value of $\Gamma_1 \approx 1.44$ (Eq.~\ref{eq:kFactor} shows that $k\propto 1/(\Gamma_1-1)$).

After ${\sim}\SI{1}{\mega \years}$ in the detached phase, $\Qshock$ peaks\footnote{
It seems counter-intuitive at first that $\Qshock$ peaks this late when $k$ and also $\dot{M}_\mathrm{gas}$ peak very shortly after detaching. However, the early extended radius keeps \Laccmax low.}, %
mainly due to a decreasing $\dot{M}_\mathrm{gas}$, and shortly after also \Ltot. As we verified separately (not shown), a high gas accretion rate tends to increase the contraction rate of the planet. Thus, it is no surprise that the decrease in \Ltot, the main contributor of which is $L_\mathrm{cntrctn}$ in a cold-start scenario, occurs shortly thereafter. 
The planet interior and surface still heat up from this point on until the electron degeneracy in the core is high enough to induce cooling (see \cref{sec:hayashiTracks}). From that point onward, most of the contraction energy $E_\mathrm{g}$ is used to raise the energy level of the degenerate electrons; they do not contribute to the internal energy of the ions nor, therefore, to the temperature.

\subsubsection{The descending branch}
\label{sec:descendingBranch}

Approximately \SI{2}{\mega \years} after detaching, the planet is past the maximum luminosity and surface temperature. This is when the planet reaches the descending branch on the HRD. With the ever vanishing disk, the gas accretion rate is constantly decreasing until the point of full disk dispersal at $\SI{9.2}{\mega\years}$. From this point on when contraction is the envelopes only energy source, $L = L_\mathrm{cntrctn}$ and the timescales $\tau_\mathrm{cool}$ and $\tau_\mathrm{KH}$ finally conjoin. 

The inflow of grains ceases simultaneously with the gas accretion (see \cref{sec:evolutionPhase}). We assume that grains quickly rain out of the atmosphere \citep{podolak_smallGrainsOpacity_2003}, so that after gas accretion has stopped, only the molecular opacities from \citet{freedman_gaseousMeanOpacities_2014} at solar composition remain. The consequent decrease in opacity in the outer envelope results in a certain increase in $L$ and $T$ on the HRD at the very end of the detached phase (this is barely visible in \cref{fig:defaultCasePlots}, but clear in \cref{fig:etaCompHRD} or \cref{fig:HRD_Lumi_twoPop}, top right). We have not investigated in detail the consequences of the exact grain opacity reduction process or the emergence of clouds but they should not modify the qualitative picture.

The rest of the planet's lifetime on the HRD is spent in the evolutionary phase in the decreasing branch, where it further contracts and cools at constant mass.
Here too, while relevant when interpreting sufficiently precise measurements (e.g., \citealp{saumon_evolution_2008,hinkley_direct_2023,morley_sonora_2024}), the microphysics of clouds does not change qualitatively the cooling of gas giants.
Contraction stagnates at around ${\sim}1\,\Rjup$ because electron degeneracy pressure prevents significant further compression (e.g., \citealp{zapolsky_mass_1969}). The resulting slope on the HRD is simply given by $L \propto T^4$ from \cref{eq:StefanBoltzmannLaw} at near constant radius. This holds as long as $T_\mathrm{int} \gg T_\mathrm{eq}$. Past $\sim\SI{4}{\giga \years}$, the planet has cooled down enough for the stellar irradiation ($T_\mathrm{eq}$ in Eq.~\ref{eq:Tsurf-evolution}) to become relevant. Thus, the \LT track starts to deviate again from the simple $T\propto L^{1/4}$ slope. Instead, the track is determined mostly by the luminosity evolution of the star \citep[e.g.][]{guillot_radiative_2010}.

\subsection{Comparing to classical cold and hot start scenarios}
\label{sec:compColdHotStart}

\begin{figure}
    \centering
    \includegraphics[width=1.0\hsize]{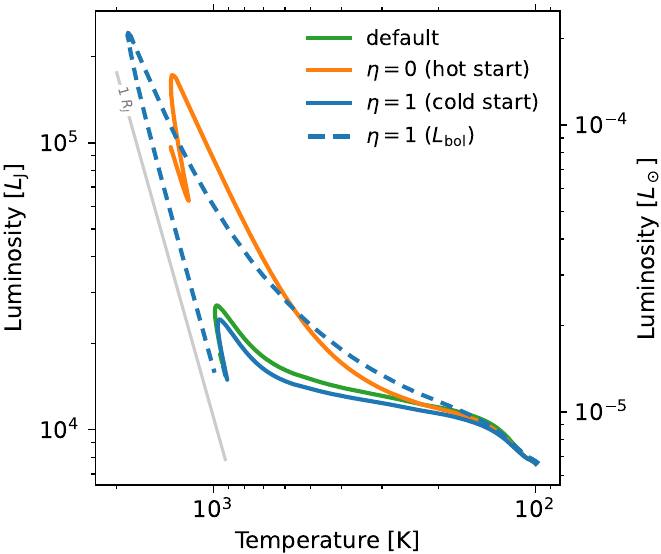}
   \caption{
    The planetary horizontal branch up to the begin of the pure cooling (evolution) phase for the default, variable-$\eta$ (green; as in \cref{fig:defHRD}), cold- (solid blue), and hot- (orange) start scenarios. Additionally, the dashed blue line shows the unobscured bolometric luminosity \Lbol an observer would see for a cold-start-accretion scenario ($\Lbol = \Lint + \Laccmax$; \cref{eq:lumi_bolo} with $k=0$), neglecting extinction.
   }
    \label{fig:etaCompHRD}
\end{figure}

Classical cold-start models \citep{marley_luminosity_2007} are not compatible with most direct-imaging observations, such as young planets in the $\beta$ Pic Moving Group \citep{gratton_implications_2024}. Thus, we compare our default intermediate scenario with fully hot and cold start scenarios. To explore those cases, we replace the factor $(k-1)$ in \cref{eq:Ltot} with a fixed value of $\eta = 1$ (cold start) or $\eta = 0$ (hot start), so that
\begin{equation}
\Ltot = L_\mathrm{int} - \eta \Laccmax.
\end{equation}
Only the resulting planetary horizontal branches are shown in the HR diagram in \cref{fig:etaCompHRD} since the ascending branch remains unaffected by our choice of $\eta$, as this parameter is only relevant during the detached phase when accreted gas impacts the accretion shock at the planet's surface.

In the horizontal branch, the hot start scenario leads to hotter and more luminous planet surfaces, reaching a maximum luminosity of $\num{1.49E-4}\,\Lsol$ in contrast to $\num{2.35E-5}\,\Lsol$ in the default and $\num{2.10E-5}\,\Lsol$ in the cold case. However, distinguishing between cold and hot starts based on bolometric luminosity only would be challenging because an observer would still see the intrinsic and radiated shock luminosity simultaneously. Separating the shock contribution (seen for example as H$\alpha$ or a UV continuum; \citealp{aoyama_spectral_2020,zhou_hubblePDS70b_2021}) from the normal photospheric emission allows to disentangle the contributions. Interestingly, in the cold-start scenario, this sum even surpasses the luminosity of the hot start case, despite identical gas accretion rates, reaching a maximum luminosity of $\num{2.09E-4}\,\Lsol$.
This occurs because the inflated radius of the hot planet results in a lower $\Laccmax$, with the energy difference being radiated away later as $L_\mathrm{cntrctn}$. Ultimately, the total integrated radiated energy is the same for both scenarios at the end of the simulation, apart from differences caused by planetesimal accretion in the detached phase due to variations in planetary radius.

Our findings on the $k$-factor from \cref{fig:kFactorOverTime} link to the early detached phase, where, for a brief period, the approach with $\Qshock$ (\cref{eq:kFactor}) can lead to more luminous planets because $k$ exceeds unity.
However, as accretion progresses, \cref{fig:etaCompHRD} shows that our default approach resembles rather a cold-start scenario than a hot one. 
At $\SI{4.2}{\mega \years}$ after detachment, the planet in our default scenario has ended its formation and its (post-formation) luminosity is $\num{1.49E-5}\,\Lsol$. This lies between the cold- ($\num{7E-6} \, \Lsol$) and hot-start ($\num{1.6E-5} \, \Lsol$) luminosities of \citet[][their Fig.~3]{marley_luminosity_2007} for a planet with comparable mass after \SI{4.2}{\mega \years}. The difference arises from the fact that we have a ${\sim}40\%$ more massive core, and that our accretion is not fully cold.

During most of the gas accretion phase, the luminosity of the hot-start planet in \cref{eq:kFactor} is primarily governed by $L_\mathrm{acc,\,max}$, with $L_\mathrm{cntrctn}$ becoming significant only once accretion ceases. In the evolutionary stage, the hot start planet retains a radius about 25\% larger than the cold-start case despite similar masses, leading to a higher luminosity due to its faster contraction. The luminosity difference between the cases diminishes over time: by \SI{1.3}{\mega \years} after formation, the hot planet is only 50\% more luminous than in the default case, declining to 10\% by \SI{11.5}{\mega \years} after formation.

\subsection{Evolution for contracting, non-accreting planets}
\label{sec:hayashiTracks}

\begin{figure}

    \includegraphics[width=1.0\hsize]{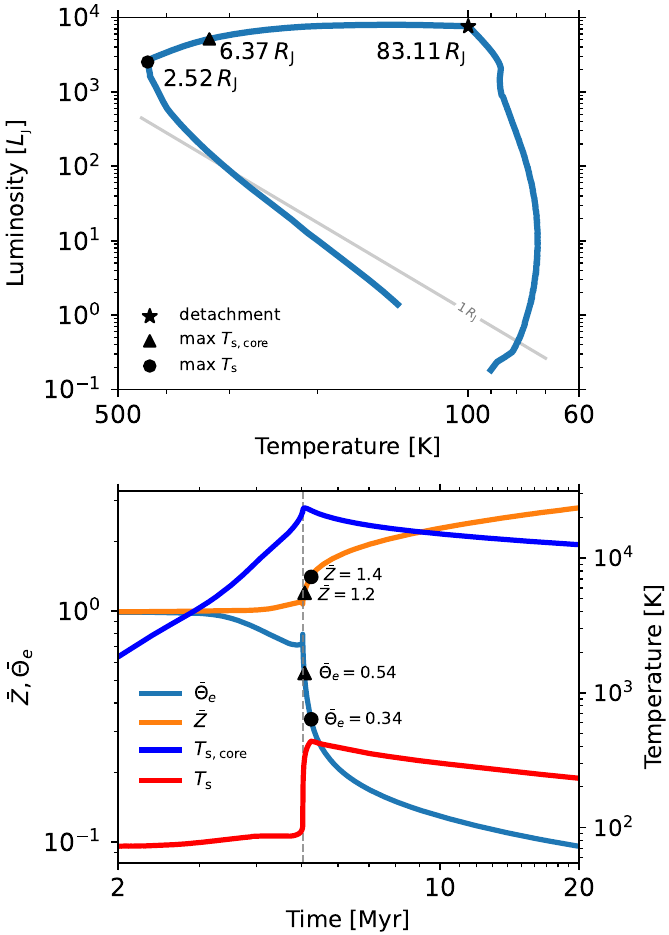}

    \caption{\textit{Top panel}: \LT track of a planet with forced total stop of accretion immediately after detaching with $M_\mathrm{tot} = 39.55\,\Mearth$, $M_\mathrm{env} = \num{21.90}\,\Mearth$ and a radius of $\num{83.11}\,{\Rjup}$. The point of detachment and end of accretion is marked with a star. The point with maximal $T_\mathrm{s,\,core} = \SI{23500}{\kelvin}$ in the forced evolution stage is marked with a triangle, while the point with $T_\mathrm{s,\,max} = \SI{436}{\kelvin}$ is marked with a dot. \textit{Bottom panel:} $\bar{Z}$ and $\bar{\Theta}_\mathrm{e}$ during the formation and forced evolution phase. The point of detachment is marked by a dashed line, and again the point of maximum \Tsurf ($T_\mathrm{s,\,core}$) is marked with a dot (triangle). Additionally, the surface temperature \Tsurf and the envelope temperature above the core $T_\mathrm{s,\,core}$ are shown in red and blue against the right vertical axis.}
    \label{fig:contractionHRD}

\end{figure}

\cite{hayashi_stellar_1961} studied in a pioneering way the development of pre-main-sequence (PMS) stars on the HR diagram, now referred to as ``Hayashi tracks''. Following rapid contraction and accretion during the protostar phase, a PMS or T Tauri star undergoes a slow contraction phase, adhering to the Hayashi tracks until nuclear fusion turns on, marking the star's reaching of the main sequence (MS). For fully convective PMS stars (with $M_\star\lesssim \SI{0.5}{\Msol}$), the (downward) Hayashi tracks are almost vertical lines on the HRD. In contrast, the tracks of more massive stars ($M_\star \gtrsim \SI{0.5}{\Msol}$) curve into vertical Henyey lines because of the emergence of inner radiative zones \citep{henyey_early_1955}.

The formation process of giant gas planets resembles that of young proto- and PMS-stars, namely in the detached phase. Immediately after detachment, the planet undergoes rapid contraction from an initial radius $\Rpla = \RBondi \sim 100\,\Rjup$ to a final radius $\Rpla \sim 2\,\Rjup$ in a short period. This phase is also when gas accretion $\dot{M}_\mathrm{gas}$ is at its peak, allowing the planet to acquire a considerable amount of its mass. This period post-detachment is akin to the protostellar scenario, which, despite rapid collapse, continues to accrete substantial mass from its surroundings.

Even though contraction in the protostellar and planetary cases might seem similar at first, there are two significant distinctions. Firstly, a planet possesses significantly less mass compared to a protostar, and thus, once accretion has stopped, will never enter a burning phase if its mass is below the deuterium burning limit. Secondly, the presence of a (rocky) core in planets leads to cooling effects and HR tracks that are distinct from classical vertical Hayashi tracks of assumed ideal core-less stars.

The results in the appendices of \citet{molliere_deuterium_2012} show the effect of a core and electron degeneracy on the cooling of the envelope. We test the applicability of those scenarios by measuring the mass-averaged compressibility factor $\bar{Z}$ and electron degeneracy $\bar{\Theta}_\mathrm{e}$. The compressibility factor $Z$ is given for each layer $i$ in the envelope by 
\begin{equation}
    Z_i = \frac{P \mu \amu}{\rho \kB T},
\end{equation}
where $\mu$ is the local mean molecular mass.
The mass average
\begin{align}
    \label{eq:ZmassAverage}
    \bar{Z} = \frac{1}{ M_\mathrm{env} }\sum_i Z_i m_i
\end{align}
measures how much the envelope resembles an ideal\footnote{%
An \textit{ideal} gas; not necessarily a \textit{perfect} gas, for which $\mu$ is constant.} 
gas, for which $Z=1$ by definition.
For $\Theta_\mathrm{e}$ we follow the implementation of \citet{molliere_deuterium_2012} and \citet{dewitt_screening_1973}: analogously to \cref{eq:ZmassAverage}, we calculate the mass average for $\Theta_\mathrm{e}$, providing a measure of electron degeneracy for our envelope ranging from $\bar{\Theta}_\mathrm{e} = 1$ (non-degenerate) to 0 (fully degenerate).

We explore $\bar{Z}$ and $\bar{\Theta}_\mathrm{e}$ in \cref{fig:contractionHRD} where we artificially turn off the accretion of gas and solids shortly after going into the detached phase in order to investigate the effect of contraction alone. The early cut-off of gas accretion leads to an unrealistically high $M_\mathrm{c}/M_\mathrm{env}$ ratio for a giant planet of ${\sim} 0.8$. Nonetheless, it enables us to study the (pure) contraction of a constant-mass envelope containing a core. According to \citet[their Appendices~A and~B]{molliere_deuterium_2012}, we anticipate that the planet's interior envelope will initially heat up until either a sufficiently small $\Rpla/R_\mathrm{c}$ ratio is attained or a notably high degeneracy $\bar{\Theta}_\mathrm{e}$ is achieved. We expect heat from contraction to be generated in rather deep layers \citep{kippenhahn_stellar_2013} but to be transported outward quickly by convection.

The test planet used in \cref{fig:contractionHRD} detaches with a mass of $M_\mathrm{tot} = \num{39.55}\,\Mearth$ and $M_\mathrm{env} = \num{21.90}\,\Mearth$ with an initial radius of $\num{83.11}\,\Rjup$. The envelope's central temperature $T_\mathrm{s,\,core}$ initially rises as the planet contracts to a radius of $\num{6.37}\,{\Rjup}$, whereas the surface temperature continues to increase until the planetary radius has decreased to $\num{2.52}\,{\Rjup}$. The lower panel of \cref{fig:contractionHRD} shows that by the time of maximum $T_\mathrm{s,\,core}$, the envelope is already significantly degenerate ($\bar{\Theta}_\mathrm{e}=0.54$), which we interpret to be the main reason for the onset of the decrease of the central temperature. 
This decrease agrees with \citet[their Appendix~B]{molliere_deuterium_2012} and is further supported by degenerate matter evolution in white dwarfs \citep{kippenhahn_stellar_2013}.
It differs from the often-invoked virial theorem for ideal (non-degenerate) gases, which assumes 
\begin{equation}
    L = -\frac{\dot{E}_\mathrm{g}}{2} = \dot{E}_\mathrm{i}.
\end{equation}
Here, a decrease in gravitational energy $\dot{E}_g<0$ suggests a rise in total internal energy $\dot{E}_\mathrm{i}$, resulting in a temperature rise for an ideal gas model. This heating is seen in the early part of the forced evolution phase, where the envelope still resembles an ideal gas and electron degeneracy is low. 

\begin{figure*}
\centering
\includegraphics[width=0.9\hsize]{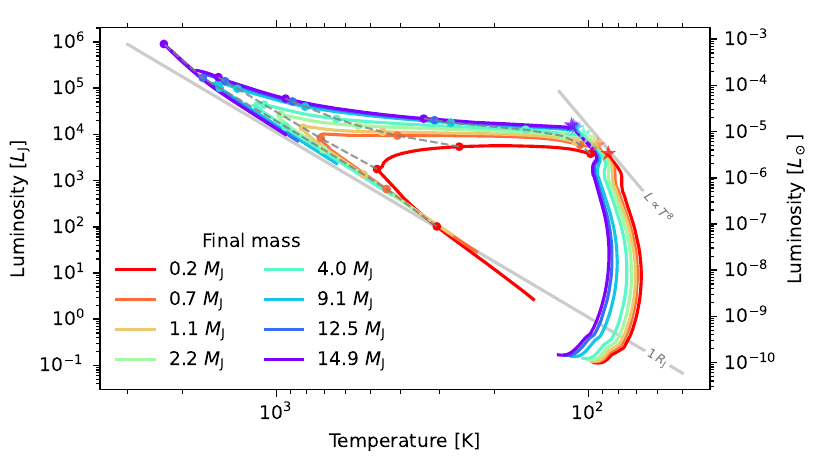}
   \caption{
    HRD tracks for planets with a range of final  masses. The different  masses are obtained by changing the initial disk gas mass at a fixed solid surface density $\Sigma_\mathrm{pla, \SI{5.2}{\mathrm{\astronomicalunit}}} = \SI{10}{\g \per \square \cm}$. The initial disk masses of \SIlist[list-units=single]{0.025;0.03;0.035;0.04;0.05;0.075;0.1}{\Msol} lead to the final masses colour-coded in the figure in the same increasing order. 
    Stars indicate the moment of detachment, while dashed lines connect isochrones at ${10^4}, {10^5}, {10^6}$ and ${10^7}$ yr after detachment.
    The most massive planet exceeds the deuterium burning limit, resulting in an additional spike in the \LT track on the top left tip. Otherwise, all masses share a similar overall shape of the HRD track with the three distinct branches. The 2.2 $\Mjup$ planet is the default case of Sect.~\ref{sec:defaultCase}. The tracks shown here end at an age of 200 Myr. Further evolution would occur approximately along the 1-$\Rjup$ line.  
    }
    \label{fig:HRD_higherMasses}
\end{figure*}

Because of this change of regime, the \LT track for this contracting planet clearly deviates from a perfectly vertical Hayashi track. \citet{guillot_giant_1996}, when investigating the detection of 51~Peg~b,  obtained a similar result, as \citet{baraffe_evolutionary_2002} did for the evolution of low-mass substellar objects (1--$20\,\Mjup$).
\citet{baraffe_evolutionary_2002} found that if the evolution begins with a large radius, \Teff increases for the first $\si{\mega \years}$ before decreasing later on. Our non-ideal gas case resembles heavily the theoretical case shown in \citet{guillot_giant_1996} (or the updated version in \citealt{guillot_giant_2023}), in which the contracting planet is avoiding the forbidden superadiabatic region. 

For the default-case planet, which has an additional energy input beyond gravitational contraction alone due to gas accretion, the maximum surface temperature is reached at $\bar{\Theta}_\mathrm{e} = 0.21$. Subsequently, the maximum central temperature of the envelope is reached at $\bar{\Theta}_\mathrm{e} = 0.18$ at a radius of $\Rpla = 1.44 \, \Rjup$.
According to \citet{guillot_interiors_2005}, the current degeneracy parameter for Jupiter is within the range of $0.03 < \theta_\mathrm{e} < 0.1$, with $\theta_\mathrm{e} = T/T_\mathrm{F}$ where $T_\mathrm{F}$ is the electron Fermi temperature. Our simulations are consistent with this, producing a value of $\theta_\mathrm{e} = 0.059$ (or $\bar{\Theta}_\mathrm{e} = 0.034$) for the default-case planet after \SI{4.6}{\giga \years}.

\subsection{Different final masses}
\label{sec:higherMassesComparison}

In \cref{fig:HRD_higherMasses}, we present \LT tracks corresponding to planets of a range of final total masses $\Mpla=0.2$--15~\Mjup. The difference in mass arises from adjustments in the initial total disk mass $M_\mathrm{disk,0}$ (\SIrange{0.025}{0.1}{\Msol}) while keeping the properties of the solids disk fixed at the same value as in the default case from \cref{sec:defaultCase}. This makes it possible to obtain comparable core masses (19--$\SI{26}{\Mearth}$) across a broad spectrum of total final masses. The values are not strictly identical because of the following reasons:
The presence of more gas in the disk leads to a higher damping effect of the planetesimal inclination $i$ and eccentricity $e$ by the gas disk and thus to a higher solid accretion rate. This in turn leads to a higher core mass and a eventually higher gas accretion rate. Moreover, a more massive disk has a higher midplane temperature \citep{nakamoto_formationDisks_1994}, which we saw in \cref{sec:defaultCase} is the main contributor to the surface temperature of the planet in the early stages of the attached phase. Altogether, in the HRD these dependences let lower-mass planets have cooler ascending branches and an earlier transition into the horizontal branch with a lower luminosity. 
The resulting shock conditions lead to relatively high shock efficiencies for all simulations. Thus, the difference in luminosity can be attributed to the higher gas accretion rates for more massive disks. On the one hand, a higher $\dot{M}_\mathrm{gas}$ directly leads to a higher $\Qshock$ given a similar value for $k$, and, on the other hand, a higher mass accreted in the early stages leads to a higher $L_\mathrm{cntrctn}$ later on, because more mass is contracting.

In general, the three different branches of the \LT tracks of all planets share a similar general shape. Except for the descending branch, the tracks run parallel, with the planets that end up more massive hotter and more luminous as their accretion rates are higher. More massive planets also pass into the horizontal branch at higher luminosities and temperatures because the disk-limited gas accretion rate is higher in the higher-mass disks.

An exception to the similar general shape is the most massive planet showing a distinct spike in the upper left corner of its HRD track in \cref{fig:HRD_higherMasses}. When the spike begins, the planet has accumulated a total mass of $\num{13.33}\,\Mjup$, exceeding the deuterium-burning threshold in the absence \citep{saumon_theory_1996} or presence of a solid core \citep{spiegel_deuteriumburning_2011,molliere_deuterium_2012, bodenheimer_deuterium_2013}. Consequently, the spike reflects the ignition of deuterium burning, resulting in a strong increase in luminosity and a temporary radius inflation. In our variable-$\eta$ scenario, which corresponds to a not very hot start scenario, $L_\mathrm{D\,burn} \gg \Qshock$, and deuterium burning at this stage rapidly becomes the dominant luminosity source ($L \approx L_\mathrm{D\,burn}$).

We can compare the results of \cref{fig:HRD_higherMasses} with \cref{sec:compColdHotStart}. We see that the hot-start formation of a $\num{2.2}\,\Mjup$  %
planet leads to a comparable maximum luminosity as of a $\SI{9.2}{\Mjup}$   %
planet in a cold-start formation scenario. However, the maximum temperatures reached differ by $\Delta T \approx \SI{380}{\kelvin}$.

While the tracks are mostly parallel in the ascending and horizontal branch, in the evolutionary descending branch all but the lowest-mass planet (0.2\,\Mjup) nearly overlap. All planets end up with approximately the same slope around the ${\simeq}\num{1}\,\Rjup$ line. This degeneracy is well known \citep{baraffe_evolutionary_2002} and results from the weak mass dependency of the radius and its weak evolution (at later times) with time for sub-stellar mass bodies with $M \sim 1$--$\num{10}\,\Mjup$ \citep{chabrier_theory_2000}. When our simulations end after $\SI{0.2}{\giga \years}$, giant planets evolve with only weakly contracting radius. Thus, the slope in the HRD follows simply from the $L \propto T^{4}$ relation. The different slope of the lowest mass planet shows that this planet with a lesser degenerate interior undergoes a more significant change in radius.

\subsection{Core-accretion scenario with pebbles}
\label{sec:twoPopComparison}

\begin{figure*}
\centering

\includegraphics[width=0.95\hsize]{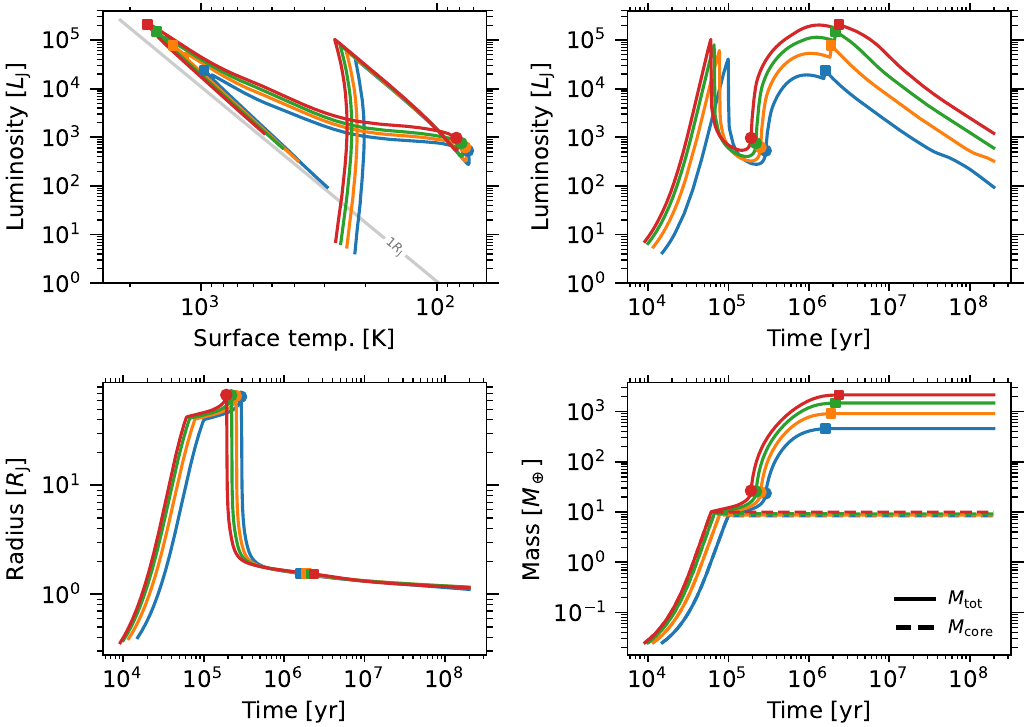}

   \caption{\textit{Top left panel}: 
    HRD for a formation scenario where the solid core grows via pebble accretion. Initial gas disk masses are 0.02~(blue), 0.03~(orange), 0.04~(green) and \SI{0.05}{\Msol} (red). The planets reach a total mass of respectively 1.4, 2.8, 4.6 and 6.7\,$\Mjup$, all with similar core masses of 8--10\,$\Mearth$. We show data up to \SI{200}{\mega\years}.
    \textit{Other panels}: Time evolution of the luminosity (without the accretion shock luminosity), radius and core and envelope masses.
    In all panels, we mark the transition from the attached to the detached phase (dot) and the time of disk dispersal (square). 
   }
    \label{fig:HRD_Lumi_twoPop}

\end{figure*}

While the previous sections focussed on planet formation simulations based on the accretion of planetesimals (10\,km in size; Table~\ref{tab:modelSetup}), which yields qualitatively similar \LT tracks, alternative accretion mechanisms can produce distinctly different formation histories. These differences arise from changes in the growth timescale and in the relative timing of solid and gas accretion, and consequently affect the thermal evolution of forming planets as traced on the HR diagram. One such scenario is pebble accretion \citep{ormel_smallBodies_2010, lambrechts_rapid_2012, ormel_emerging_2017}, where typically cm-sized particles are the primary solid building blocks. This mechanism can facilitate rapid core growth at large distances.

We present \LT tracks for simulations where solid accretion occurs only via pebbles using an alternative solids disc model, namely the two-population model of \citet{birnstiel_simple_2012}. The general implementation of this model within the Bern Model is covered in \citet{voelkel_pebbleFlux_2020} and \citet{emsenhuber_toward_2023}. We use the slightly adapted prescription of \citet{ormel_emerging_2017} for the accretion of pebbles, with a sharp cut-off at the pebble isolation mass \citep{bitsch_pebble-isolation_2018}. To create planets with masses comparable to those in previous sections, we use total disk masses of $M_\mathrm{disk,0} =0.02$, 0.03, 0.04, \SI{0.05}{\Msol} and an external photo-evaporation rate of $\dot{M}_\mathrm{wind} = \SI{E-5}{\Msol \, yr^{-1}}$. 

\cref{fig:HRD_Lumi_twoPop} shows the \LT tracks of the four resulting planets, along with the time evolution of their luminosity (without the accretion shock luminosity), radius, and mass. Unsurprisingly, the most notable difference occurs during the attached phase, where the ascending branch splits into two distinct parts. In the early stages, the solids accretion timescale $\tau_\mathrm{acc,s}$ is approximately two orders of magnitude shorter than in the planetesimal accretion scenario, resulting in substantially higher total luminosities at early times.
During this phase, the planets rapidly ascend the initial, steep (approximately vertical) branch until pebble accretion ceases once the pebble isolation mass of about 9 $\Mearth$ is reached, after roughly 60--100~kyr, depending on the initial disk mass. The pebble isolation mass is not identical, as the scale height, which is a factor setting the pebble isolation mass, depends to some extend on the disk mass.

Once the pebble isolation mass is reached, the planetary luminosity is at a local maximum. For the planets with a lower final total mass, this is even a global maximum. Because of this hot and luminous state, gas accretion remains low and insufficient to let the planet detach from the nebula. This effect of the suppression of gas accretion because of the high solid accretion rate with pebbles was studied in detail in \citet{kessler_interplay_2023}. 

Next, with the abrupt termination of the pebble flux, the core luminosity, and hence $\Ltot$, decreases rapidly.
This subsequent cooling phase lasts for approximately \SI{0.1}{\mega\years}. During this time, the mass is approximately constant and still almost equal to the core mass. 
The total radius \Rout is already dominated by the Hill radius term ($R_{\mathrm{H}}/4\ll\RBondi$), so that \Rout,
which scales with the Hill sphere, is also constant.
Therefore, the tracks next move diagonally downward to smaller temperature and luminosity with $L\propto T^4$. The temperature decreases in this phase from 200--300~K to about 70--80~K which is only slightly more than the disk background temperature. The tracks are thus approximately parallel to the gray line in the top left panel representative of the slope of the tracks in the evolutionary phase (descending branch) where the radius is also varying little, but of course in the present phase the radius is much larger. 

This diagonal downward movement is halted once gas accretion regulated by the classical Kelvin--Helmholtz cooling of the gaseous envelope at zero solid accretion \citep{ikoma_formation_2000} becomes significant. Because of gas accretion the luminosity begins to rise again. The KH-regulated gas accretion rate increases as runaway gas accretion commences and soon exceeds the disk-limited value, leading to detachment which happens at about 20--30\,$\Mearth$ and about \SI{0.1}{\mega\years} after the luminosity minimum.

\begin{figure*}
    \begin{minipage}[c]{0.5\textwidth}
    \centering
        \includegraphics[width=95.5mm]{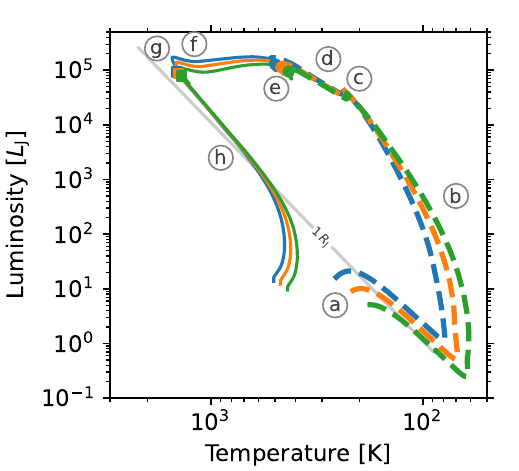}
    \end{minipage}%
    \hfill
\begin{minipage}{88mm} 
    \centering
    \renewcommand{\arraystretch}{1.5}
        \begin{tabularx}{88mm}{X}
            \mycirc{a} Embryo mass reaches $2 \, M_{\leftmoon}$ \\
            \mycirc{b} $t{\approx}\SI{E5}{\years}$: $\dot{M} {=} \dot{M}_\mathrm{s}$, slow Type~I migration with $\tau_\mathrm{mig} > \tau_\mathrm{acc}$ \\
            \mycirc{c} Fast inward migration in Type~I regime with $\tau_\mathrm{mig} < \tau_\mathrm{acc}$ and thus reduction of $R{\approx}R_\mathrm{H}$. Sudden drop of $\Sigma_\mathrm{pla}$ at ice line \\
            \mycirc{d} Type~I migration to inner disk and transition into Type~II \\
            \mycirc{e} Switch to slower Type~II migration. $L_\mathrm{M,s}$ drops as planet clears its feeding zone before $L_\mathrm{M,g}$ starts to rise before $M_\mathrm{c}{=}M_\mathrm{env}$ and then detaching at $\bullet$ \\
            \mycirc{f} Horizontal branch: rapid radius contraction phase with almost ideal gas like envelope \\
            \mycirc{g} Start of significant degeneracy in interior and envelope cooling, slower radius contraction. Disk dispersal later at $\blacksquare$ \\
            \mycirc{h} Long term cooling and contraction at constant mass and near constant radius
        \end{tabularx}
\end{minipage}

\includegraphics[width=1\hsize]{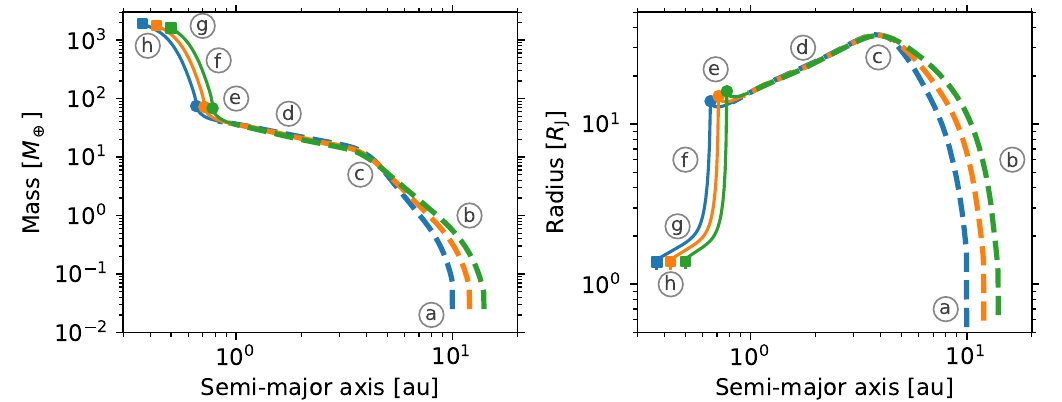}

   \caption{\LT tracks (\textit{top panel}) of three migrating planets with initial semi-major axis of \SI{10}{\astronomicalunit} (blue), \SI{12}{\astronomicalunit} (orange) and \SI{14}{\astronomicalunit} (green) showing their formation and evolution up until \SI{200}{\mega\years}. Filled dots and squares show the point of detachment and disk dispersal, respectively. Labels \marker{a}--\marker{h} mark various phases in the process, explained in the top right.
   Type~I (Type~II) migration is indicated by dashed (solid) segments.
   \textit{Bottom panels:}
   Mass and radius as a function of orbital distance while the planets migrate inwards. Final masses are 5--6~\Mjup.
   }
    \label{fig:migration_HRD_am-R}
\end{figure*}

\begin{figure*}
\centering

\includegraphics[width=0.9\hsize]{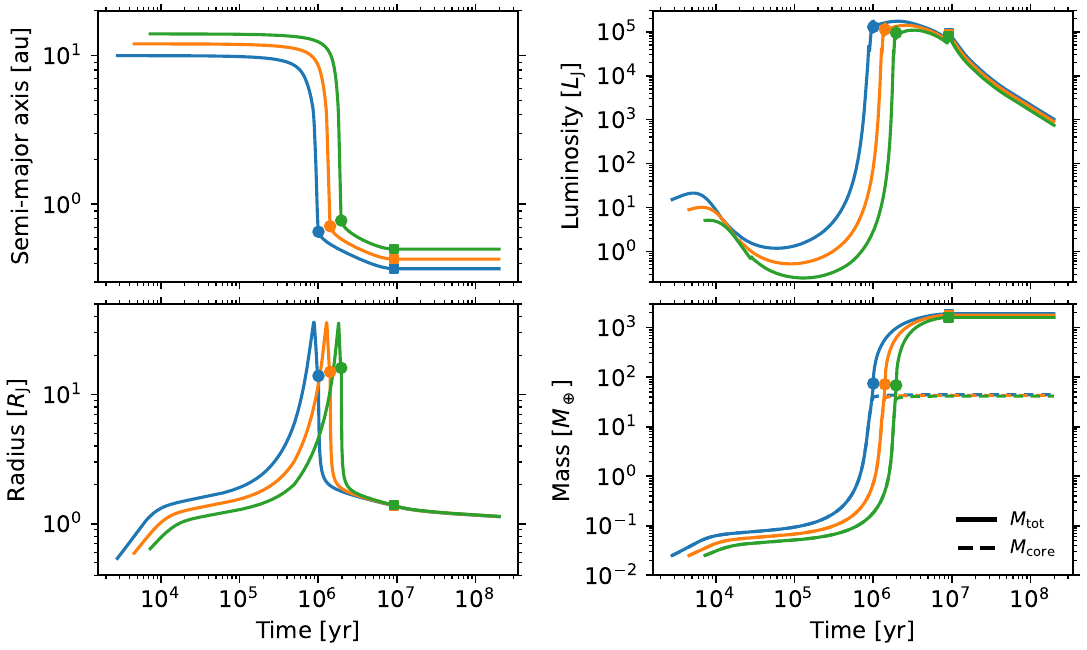}

   \caption{
    Evolution of the semi-major axis, luminosity, radius and total mass for three migrating planets with different initial starting location of \SI{10}{\astronomicalunit} (blue), \SI{12}{\astronomicalunit} (orange) and \SI{14}{\astronomicalunit} (green). The point of detachment is marked by a solid dot and the point of disk dispersal by a solid square. Data is shown for the formation phases and until \SI{200}{\mega\years} of the evolution phase and not for the full duration as previously in \cref{fig:migration_HRD_am-R}.
   }
    \label{fig:migratingPlanets-multiPlot}
\end{figure*}

Thus, the different solid accretion models lead to clearly different tracks early on. We note en passant that the HR tracks would likely have a similar shape for (hypothetical) runaway planetesimal accretion (instead of oligarchic) as assumed in the work of \citet{pollack_formation_1996}, but we have not run such simulations directly.

Remarkably, the early luminosity peak caused by pebble accretion may reach values comparable to, or in some cases even exceeding, the maximum luminosity attained during the detached phase (rapid gas accretion), placing it for scale in the same order of magnitude as the inferred luminosity of directly imaged forming planets such as PDS70~b \citep{stolker_miracles_2020,wang_constraining_2021,blakely_james_2025}. However, at this stage the planet is still deeply embedded in the protoplanetary disk, and absorption by the surrounding material likely hinders significantly its detectability. Additionally, it is a very short phase lasting only some 10~kyr.

After the initial ascending branch and detachment from the disk, the subsequent planetary horizontal branch and the descending branch during evolution closely resemble those obtained in the planetesimal accretion scenario.

\subsection{Migrating planets}
\label{sec:migratingPlanets}

In the previous sections, orbital migration was neglected in order to simplify the processes governing the formation and thermal evolution of young planets. Here, we include migration and examine its impact on the resulting \LT tracks. To this end, we consider single-embryo formation simulations with initial semimajor axes of 10, 12, and 14~\si{\astronomicalunit}, producing planets with total masses between 5~and 6~\Mjup.
Also, we adopt the gas accretion prescription of \citet{bodenheimer_deuterium_2013} and, to ensure the formation of giant planets also at larger distances, reduce the planetesimal size to \SI{1}{\km}. Finally, we replace the
constant-powerlaw solids disk profile (Eq.~\ref{eq:Sigmasimal}) used up to now
with a profile that follows the gas disk, scaling the gas surface density by a constant dust-to-gas ratio $\fpg=0.03$.

\cref{fig:migration_HRD_am-R} presents the resulting HRD of the migrating planets, together with the evolution of their mass and radius as a function of semimajor axis. The three panels use the same markers \marker{a}--\marker{h} to indicate key phases and events, as annotated in the figure. To enable a comparison of the respective timescales, \cref{fig:migratingPlanets-multiPlot} shows the evolution of $a$, $L$, $\Rpla$, and $\Mpla$ over the first \SI{200}{\mega\years}.

A short episode of enhanced planetesimal accretion during the first \SI{10000}{\years}, occurring just before the planet reaches a mass of \num{0.025}~$\Mearth$ (${\sim}2$ lunar masses) at \marker{a}, produces the prominent feature on the early ``ascending branch'' of the HRD. This behaviour is not specific to these runs with migration; in the in-situ simulations shown before, the same phase occurs at lower masses not shown owing to the different disk setup and initialisation. 
The ascending branch around \marker{b} is similar to that in the in-situ case. The luminosity is dominated by solids accretion ($L = L_\mathrm{M,s}$), while the surface temperature remains influenced by the disk midplane temperature.

As the planet grows, however, the slope of the \LT track increasingly deviates from the steep ascending branches found in the in-situ runs. This deviation arises from the migration of the planet within the disk, which modifies both the local disk conditions and the solids accretion rate. Young, low-mass planets embedded in protoplanetary disks initially undergo inward Type~I migration (dashed line) \citep{goldreich_excitation_1979,ward_protoplanetMigration_1997,papaloizou_disk-planet_2007}. 
Type~I migration can occur at remarkably high rates, as shown in \cref{fig:migratingPlanets-multiPlot}.
For example, the planet initially situated at \SI{14}{\astronomicalunit} migrates inward from \SI{10}{\astronomicalunit} to \SI{1}{\astronomicalunit} within just ${\sim}\SI{0.5}{\mega\years}$.
As the planets sweep through the disk, there is a sudden reduction of $\Sigma_\mathrm{pla}$, and thus $\dot{M}_\mathrm{s}$, at the ice-line, which causes a break in the luminosity (marker \marker{c} in the HRD in \cref{fig:migration_HRD_am-R}). The accretion rate quickly rebounds as the planets keep migrating toward the inner disk, where planetesimal surface densities are once again higher.

Another notable consequence of migration is the reduction of the planetary radius during the attached phase, driven by the linear dependence of the Hill radius $R_\mathrm{H}$ on the semi-major axis, despite the concurrent increase in planetary mass, which enters the Hill radius only weakly, scaling as $\propto \Mpla^{1/3}$.
Compared to the previous in-situ simulations studied above, the simulations incorporating migration yield higher core masses of ${\sim}\num{40}~\Mearth$ \citep{ward_rapid_1989,alibert_migrationAndFormation_2004}, resulting in larger luminosities at the end of the ascending branch. In combination with the reduced planetary radii, this leads to higher radiative fluxes and hence to significantly higher intrinsic and surface temperatures (see Eq.~\ref{eq:Tint4}). In addition, inward migration places the planet in progressively hotter regions of the protoplanetary disk, further increasing the surface temperature. As a result, during the early attached phase, migrating planets follow ascending branches in the HR diagram that are shifted toward higher temperatures and inclined to the left compared to the in-situ case. The resulting evolutionary tracks therefore also deviate from the previously observed $L\propto T^8$ relation, as migration is most rapid precisely when the approximation $R \approx R_\mathrm{H}$ becomes valid.

Around \marker{e}, three key transitions occur in close succession: the planet's envelope mass becomes comparable to the core mass, $M_{\rm env}\simeq M_{\rm c}$, the planet detaches from the disk (end of the attached phase), and migration slows from Type~I to Type~II as the planet begins to open a gap (\citealp{papaloizou_formationAndMigration_2006}; Section~5.1 of
\citealp{emsenhuber_ngpps1_2021} and references therein). In the present simulations we find that detachment occurs approximately when the core and envelope have the same mass (the ``crossover point''; e.g., \citealp{pollack_formation_1996,hubickyj_accretion_2005}), and that the transition from Type~I to Type~II  precedes this by only 30--40~kyr.
This near-coincidence is expected because all three events are primarily controlled by the rapid increase in planet mass: once the planet is massive enough to perturb significantly the disk (reducing the migration torque and initiating gap opening), it is also close to the onset of runaway gas accretion, causing $M_{\rm env}$ to rise rapidly and leading to detachment.

The subsequent formation and evolution from the horizontal branch on from \marker{f} to \marker{h} resembles the in-situ simulations shown earlier. At \marker{f}, the envelopes, initially still similar to an ideal gas, contract rapidly and heat up until the overall degeneracy is high enough to halt the fast shrinking. From this point on at \marker{g}, while still accreting gas, the planets contract and cool. By this stage, migration has moved the planets close to the star, where the irradiation slows their cooling. At the age of the Solar System, their surface temperature is ${\gtrsim} \SI{500}{\kelvin}$.

\section{Discussion and conclusions}
\label{sec:discussion}

Our results illustrate the strong imprints of the main phases of the core-accretion paradigm -- whether core growth happens through planetesimals or pebbles -- on the luminosity and temperature evolution and thus the HRD tracks of forming planets. This impact is particularly significant in the early stages in the ascending branch. In the pebble-accretion scenario, the rapid accretion of solids leads to a brief but intense luminosity spike, with peak luminosities reaching up to ${\sim}\num{E5}\,{\Ljup}$ (${\sim}10^{-4}$~\Lsol) at ages ${\lesssim}\SI{0.1}{\mega\year}$, during which the \LT tracks show a sharp upward trend at a nearly constant temperature. In contrast, planetesimal accretion produces a more gradual ascent at lower temperatures, reflecting the longer solid accretion timescale. 
This clear separation of formation pathways in the HR diagram highlights the diagnostic power of this representation, as differences in the underlying accretion physics translate directly into distinct tracks.

\begin{figure*}[t]
  \centering

    \hspace*{\fill}
  \includegraphics[width=0.45\linewidth]{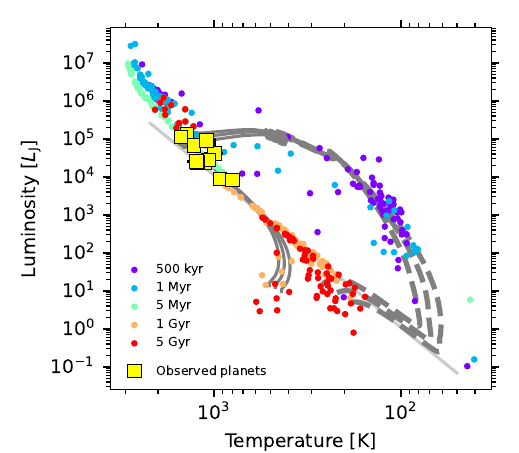}
  \hspace*{\fill}
  \includegraphics[width=0.45\linewidth]{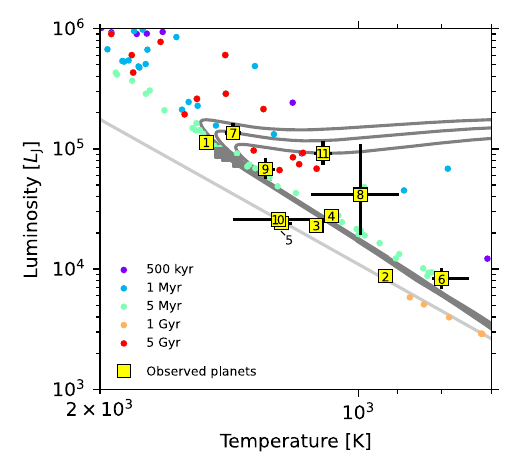}
  \hspace*{\fill}

  \caption[]
{\textit{Left panel}: \LT tracks for migrating planets from \cref{fig:migration_HRD_am-R} (grey lines), overlaid with directly-imaged companions from \cref{tab:data_directly_imaged_planets} (yellow squares). 
We show planets in the \texttt{NG76} hot-start population ($\eta = 0$) synthesis \citep{emsenhuber_ngpps1_2021}, taken from the Data Analysis Centre for Exoplanets (DACE)\footnotemark, at five different times (coloured dots). We display only planets with a final mass $\Mpla \geq 0.9\Mjup$. Of the 82 synthetic planets, 16 end up with a mass above the deuterium-burning limit. 
\textit{Right panel}: Zoom-in around the young directly-imaged planets. Numbering as in \cref{tab:data_directly_imaged_planets}.}
    \label{fig:DI-data}
\end{figure*}

The high luminosities reached during the early pebble accretion phase raise interesting questions regarding the observability of such young, luminous yet still embedded planets. \citet{brunngraber_constraints_2018}, for instance, predict detections via brightness asymmetries for luminosities above $L\ga \num{5E6}\,{\Ljup}$  %
(their test companion $\mathcal{E}$), measurable with the MATISSE beam combiner \citep{lopez_overview_2014}.
While the luminosities obtained in our simulations (\cref{sec:twoPopComparison}) remain below this threshold, higher pebble isolation masses, as suggested by \citet{bitsch_pebble-isolation_2018}, could potentially yield more luminous objects. A population-synthesis-based study would be required to explore this regime in a statistically meaningful way. 
In that embedded, early phase, luminosity may in fact be the main observational tracer for young, low-mass planets, as they are not expected to open a gap or generate detectable disc substructures in Class~0/I systems \citep{nazari_hidden_2025}. A high luminosity could also help make even low-mass planets detectable with the Atacama Large Millimeter/submillimeter Array (ALMA) through their indirect consequences on the disc chemistry \citep[e.g.][]{yoshida_protoplanet_2026}.

More generally, planetary HRD provide insight into the importance of the physical processes governing the  luminosity and surface temperature of a planet at different formation stages. Heating and cooling due to contraction, gas accretion, and orbital migration imprint distinct signatures on the tracks, allowing the dominant energy source to be identified from the location and slope in the HR diagram. In contrast to time-based evolutionary representations, the HR diagram makes these different contributions apparent through changes in the track shape.

Identifying the dominant luminosity sources during formation is also a prerequisite for constructing realistic spectral energy distributions of young, forming planets and for interpreting accretion-related observables. Our treatment of the accretion shock provides a starting point for a more detailed analysis of emission diagnostics such as H$\alpha$ during the formation phase. However, the impact of more complex gas accretion geometries through a circumplanetary disk (e.g. \citealt{ward_circumplanetary_2010}; \citealt{zhu_accreting_2015}; \citealt{szulagyi_observability_2019}) is not captured in the present model and will be investigated in future work.
A natural next step is to couple the thermal-evolution tracks to circumplanetary disk accretion and radiative-transfer calculations in order to predict full SEDs and observable magnitudes and colours, rather than intrinsic bolometric luminosities alone. This will enable direct comparisons in formats like colour--magnitude diagrams and will help assess how accretion geometry and reprocessing in the CPD modify the inferred effective temperatures and luminosities (e.g. \citealt{aoyama_spectral_2020,marleau_accretingProtoplanets_2022,adams_analyticApproach_2022,marleau_accretionShock-III_2023,choksi_spectral_2025,taylor_2D_2026}).
In this context, explaining the high entropies (hot-starts) inferred for some directly imaged planets (e.g. \citealt{gratton_implications_2024}) remains an open challenge unless relatively massive cores provide the necessary heating \citep{mordasini_luminosity_2013}. 

\footnotetext{The full data set can be visualised at and downloaded from the Data Analysis Centre for Exoplanets (DACE) at \url{https://dace.unige.ch}.}

In this comparison it is important to distinguish between ``bare'' young planets with weak or negligible ongoing accretion and a faint (or absent) circumplanetary disk and very young, still embedded and actively accreting planets. The \LT tracks presented in this work describe the evolution of the planet as driven mostly by its interior. These tracks are therefore most directly comparable to the former category, for which the observed emission is expected to be dominated by the planet photosphere.

For the more embedded, strongly accreting objects, however, the emergent spectrum and broadband magnitudes can be dominated by contributions from the accretion shock and the circumplanetary disk itself and/or its reprocessing of the emission from the planet, so that a direct comparison to intrinsic \LT tracks becomes less meaningful. Extending the planetary HRD concept into that regime will require forward modelling that consistently combines photospheric emission, accretion-shock luminosity, and circumplanetary disk emission into synthetic SEDs and colours, providing a pathway toward interpreting young planets in colour--magnitude space and toward a more complete observational classification across formation stages. Steps in this direction have appeared recently in the literature (\citealp{sun_observational_2024}; Sun et al., subm.).

With this limitation in mind, we compare the tracks to observed and synthetic planets.
In \cref{fig:DI-data}, we show selected data from the NASA Exoplanet Archive for directly imaged companions (yellow squares) along with population synthesis data from \citet{emsenhuber_ngpps1_2021} (coloured dots) and \LT tracks of migrating planets from this work (grey lines, as in \cref{fig:migration_HRD_am-R}). The position of the directly-imaged companions on the HRD provides a qualitative indication of their formation phase or evolutionary stage. Unsurprisingly, apart from PDS70~b/c and WISPIT~2~b (discussed next), all lie close to the $1\,\Rjup$ line characterising the descending evolutionary branch. This is expected, since the listed companions are mostly found in systems without any significant accretion disks so that accretion has already terminated or nearly so. 

\begin{table*}[t]
    \caption{List of data for directly imaged planets from the NASA Exoplanet Archive (\url{https://exoplanetarchive.ipac.caltech.edu/docs/data.html}). For WISPIT~2 b, we used the estimated mass and age from the discovery paper and looked up standard cooling tracks.}
    \label{tab:data_directly_imaged_planets}
    \centering
    \begin{tabular}{ccccl}
    \hline
    \hline
\# & Planet            & $\log(L/L_\odot)$       & \Teff $[\si{\kelvin}]$ & Source                                 \\ \hline
1  & $\beta$ Pic b 
& $-4.01^{+0.04}_{-0.05}$
& $1503\pm2$ 
  & \citet{ravet_multimodal_2025}           \\
2  & HR 8799 b 
& $-5.12\pm0.01$ 
& $930\pm15$
  & \citet{xuan_HR8799composition_2026}              \\
3  & HR 8799 c 
& $-4.70\pm0.1$
& $1120\pm15$
  & \citet{xuan_HR8799composition_2026}              \\
4  & HR 8799 d
& $-4.62\pm0.1$
& $1075\pm20$            
  & \citet{xuan_HR8799composition_2026}              \\
5  & HR 8799 e
& $-4.68\pm0.02$ 
& $1230\pm35$    
  & \citet{xuan_HR8799composition_2026}              \\
6  & 51 Eri b
& $-5.14\pm0.09$
& $800^{+22}_{-56}$                
   & \citet{madurowicz_direct_2025}            \\
7  & PDS 70 b          
& $-3.93\pm0.05$ 
& $1400\pm30$      
  & \citet{blakely_james_2025}            \\
8  & PDS 70 c         
& $-4.44^{+0.42}_{-0.33}$
& $995^{+141}_{-97}$                & \citet{wang_kecknirc2_2020}              \\
9  & HIP 65426 b     
& $-4.23\pm0.09$
& $1283^{+25}_{-31}$              & \citet{carter_jwst_2023}           \\
10 & TYC 8998-760-1 c 
& $-4.65^{+0.05}_{-0.08}$
& $1240^{+160}_{-170}$              & \citet{bohn_two_2020}                    \\
11 & WISPIT~2 b  
& $-4.1\pm0.1$
& $1100\pm30$                & \citet{vanCapelleveen_WISPIT_2025}       \\ 
    \hline
    \hline
\end{tabular}
\end{table*}

The data points from the population synthesis of \citet{emsenhuber_ngpps1_2021} accumulate in the upper-left region. A key reason for this is that the population we used, ``\texttt{NG76}'', assumes the classical hot-start scenario ($\eta = 0$), leading to hot and luminous planets even below the brown-dwarf threshold. In addition, the broad mass range of the sample (${\sim}1$--30~\Mjup) further contributes to this pattern; the most luminous objects in the upper left corner predominantly correspond to higher-mass planets and brown dwarfs (16 out of the total 82 planets shown end up with a mass above the deuterium-burning limit), while lower-mass objects occupy the fainter, cooler part of the distribution.
PDS70~b/c might eventually reach the upper left area, since a cold-start scenario seems to be ruled out for these planets \citep{trevascus_differentiatingPDS70_2025}. Otherwise, as it could also be for WISPIT~2~b, if their formation is already over and they have begun their pure cooling phase, they are starting at a slightly higher radius than other objects.

Our \LT tracks show only the planet's total intrinsic luminosity, and not the one an observer would see. Depending on the fate of the shock luminosity like absorption, this would rather be \Lbol, which includes also the luminosity emitted at the accretion shock. We have seen in \cref{sec:compColdHotStart} that the observed luminosity might be up to an order of magnitude higher than the intrinsic one for cold-start scenarios (\cref{fig:etaCompHRD}). Thus, PDS70~b's position on the HRD is to some extent also compatible with the tracks of lower-mass planets with ${\sim}\num{1}\,{\Mjup}$.

Altogether, the tracks presented here provide a framework for interpreting the luminosity--temperature evolution of forming and young giant planets in a representation analogous to the stellar HRD, a concept well known in astronomy.
Placing directly-imaged companions on the planetary HRD, together with model tracks, can help identify whether an object is plausibly still accreting, undergoing rapid post-detachment contraction, or already evolving on a near-constant-mass cooling track. At present, however, because of the difficulty of detecting deeply embedded planets, the directly imaged sample is small and biased toward relatively evolved, weakly accreting (or non-accreting) objects, so that the use of planetary HRDs for population-level inferences remains largely prospective. In this sense, the planetary HRD offers a physically motivated way to relate observed luminosities and inferred temperatures to formation histories, while keeping in mind that additional processes affecting observability (extinction and absorption) must be treated separately.

With the work presented in this paper, a well-known framework, the HR diagram, is extended onto the subject of planet formation. This provides an intuitive way to link observed luminosities and temperatures to formation physics and also lays the groundwork for future studies aiming to classify directly imaged young planets. However, it will be challenging to populate observationally the short early phases in the planetary horizontal branch. As discussed in Sect.~\ref{sec:horizontalBranch} and seen in the population-synthesis snapshots in \cref{fig:DI-data}, because the horizontal branch phase is brief, only a small fraction of objects is expected to be caught there in any given snapshot, making it intrinsically difficult to build a large observed sample in that region. In practice, detecting even earlier stages on the ascending branch will be further complicated by absorption and reprocessing within the disk and envelope surrounding the forming planet.

\section{Summary}
\label{sec:summary}

This study explored the concept of a ``planetary Hertzsprung--Russell diagram (HRD)'', expanding the traditional stellar HRD to planetary systems and including in particular the formation phase of gas giants.
Akin to the stellar case, the planetary HRD displays
the surface temperature \Tsurf against the total luminosity emerging from the surface \Ltot.
The planetary HRD provides a framework for visualising and analysing planetary formation tracks across diverse scenarios.
In this work, we used tracks computed self-consistently with a detailed 1D global giant planet formation model based on the core accretion paradigm, the Generation III Bern Model \citep{emsenhuber_ngpps1_2021}.

We emphasise that this first paper is meant to be a theoretical study. Given the known simplifications and limitations of the model used here which are of relevance especially for the link to observations -- such as the absence of a CPD or the treatment of absorption and extinction -- our main goal is not yet to make direct observational predictions. Instead, we aim to offer a theoretical context of the processes governing planet formation and their expression in a hitherto little explored way, setting the basis for future links between formation theory and current and hopefully numerous future observations of directly imaged planets, forming planets, and protoplanetary disks. Expanding on these points will be the subject of future work.

The main results can be summarised as follows:
\begin{itemize}

    \item In the scenario where the solid core forms via the accretion of planetesimals (Sect.~\ref{sec:defaultCase}), the \LT tracks show three distinct branches corresponding to the main phases in giant planet formation \citep{bodenheimer_models_2000,mordasini_characterization-I_2012}: an ascending branch during the attached phase, a planetary horizontal branch shaped by the rapid radius contraction and runaway gas accretion in the detached phase, and a descending branch during the post-formation evolution (cooling) at constant mass (Fig.~\ref{fig:defHRD}).
    
    \item In the ascending branch (Sect.~\ref{sec:ascendingBranch}), where planets move approximately vertically upward in the HR diagram, the luminosity is set primarily by the accretion of solids ($L \approx L_\mathrm{M,s}$), while \Tsurf is still significantly influenced by the temperature of the background nebula through $\Tsurf^4 = T_\mathrm{int}^4 + T_\mathrm{d}^4$. Typically, \Tsurf is low due to the extended planetary radius in the attached phase ($\Rpla \sim 100\,\Rjup$). In the late attached phase when $\Rpla\approx R_\mathrm{H}/4$ and $T_\mathrm{int} \gg T_\mathrm{d}$, we find analytically for in-situ planetesimal accretion the scaling $L\propto T^8$ (Appendix~\ref{appendix:appendixB_slopeAttachedPhase}). This explains the slope seen in the numerical simulations.
    
    \item The planetary horizontal branch (Sect.~\ref{sec:horizontalBranch}), where planets move approximately horizontally to the left, starts with the onset of disk-limited gas accretion and detachment from the disk. Detachment typically happens shortly (a few 10 kyr) after crossover ($M_{\rm c}=M_{\rm env}$) is reached and rapid gas accretion begins. Detachment shows up as a sharp knee in the \LT track. The subsequent rapid contraction from very large radii down to a few Jovian radii leads to a nearly horizontal motion: a strong increase in \Tsurf from $T_\mathrm{d}\sim 100$~K to $\sim 1000$--1500~K with a modest increase in \Ltot. This occurs over short timescales ($\sim 10^5$~yr).
    
    \item The descending branch (Sect.~\ref{sec:descendingBranch}) starts once electron degeneracy becomes important. Then, the radius becomes only weakly dependent on temperature, which in turn stops the rapid contraction. This leads to another knee in the \LT track. Further evolution occurs at nearly constant radius, which decreases from $\approx$2 to 1 $\Rjup$ over Gyr timescales. The \LT tracks thus now simply follow $L \sim T^4$; the planets move diagonally, downwards to the right in the HRD. This holds until stellar irradiation becomes relevant for the temperature at late times.
    
    \item For the first time, we have replaced the usual assumption of fixed hot or cold accretion (via a constant heating efficiency) by a time-dependent relative shock heating factor $k$ (Sect.~\ref{sec:shock_conditions}) based on radiation-hydrodynamic simulations \citep{marleau_accretionShock-I_2017, marleau_accretionShock-II-2019, marleau_accretionShock-III_2023}. We found that spherically symmetric gas accretion yields a brief warm to hot accretion period shortly after detachment but that later on, $k$ becomes small, corresponding to cold accretion, for most of the runaway gas accretion phase.
    
    \item We have studied \LT tracks for final planet masses ranging between 0.2 and 15~$\Mjup$ (Sect.~\ref{sec:higherMassesComparison}). The basic shape of ascending, horizontal and descending branches is retained in all cases. The formation of higher-mass planets leads to more luminous and hotter tracks on the HRD compared to those of their lower-mass counterparts. For sufficiently massive objects ($\Mpla\gtrsim 13\,\Mjup$), deuterium burning generates a distinctive spike of the track towards the upper left knee, near the transition from the horizontal to the descending branch (Fig.~\ref{fig:HRD_higherMasses}).
    
    \item In the scenario where the accretion of the core happens via pebbles (Sect.~\ref{sec:twoPopComparison}), the attached phase splits into two stages: an early, brief, high-luminosity phase while pebble accretion is active, followed by a cooling phase at nearly constant core mass before Kelvin--Helmholtz-regulated gas accretion triggers detachment. This changes the form of the \LT tracks significantly. The first phase in which the luminosity is powered by solid accretion is still present as a steep ascending track until the the pebble isolation mass is reached. The subsequent cooling translates into an intermediate descending cooling track with $L\propto T^4$ (with the radius equal to $R_\mathrm{H}$ and initially a weakly increasing envelope mass) until the point of detachment is reached when gas accretion reaches the disk-limited value. The ensuing planetary horizontal branch then appears with a slope towards higher luminosities than what is seen in the planetesimal case, since the absence of solids accretion leads to a less luminous starting point for the fast horizontal movement. Additionally, the cooling after the first phase means that the horizontal branch crosses the early steep ascending branch.
    
    \item We have studied the effect of orbit migration (\cref{sec:migratingPlanets}). Migration shifts the attached-phase tracks to higher \Tsurf and modifies their slopes, because the planets move into hotter disk regions and because $R\sim R_\mathrm{H}\propto a$ decreases during inward migration. 

    \item Placing observed and synthetic planets together with \LT tracks of migrating planets in the planetary HRD (\cref{fig:DI-data}) provides a qualitative indication of the formation phase or evolutionary stage of the directly imaged companions in \cref{tab:data_directly_imaged_planets}.

\end{itemize}

\begin{acknowledgements}
B.G.\ and C.M.\ acknowledge the support from the Swiss National Science Foundation under grant 200021\_204847 ``PlanetsInTime''. G.-D.M.\ acknowledges the support of the Deutsche Forschungsgemeinschaft (DFG) through grant MA~9185/2-1. Part of this work has been carried out within the framework of the NCCR PlanetS supported by the Swiss National Science Foundation under grants 51NF40\_182901 and 51NF40\_205606. Calculations were performed on the \texttt{horus} clusters of the Division of Space Research and Planetary Sciences at the  University of Bern.
\end{acknowledgements}

\bibliographystyle{aa_url}
\bibliography{beniBib}

\begin{appendix}

\section{The slope of HR tracks  in the ascending branch due to planetesimal accretion}
\label{appendix:appendixB_slopeAttachedPhase}

In this appendix we derive an approximation for the slope of the upper part of the ascending branch in the planetary HR diagram, i.e.\ the proportionality between $L$ and $T$.
During all stages, by definition of the effective temperature, we have
\begin{equation}
    \label{eq:LsigmaT4}
    L = 4 \pi R^2 \sigma T^4.
\end{equation}
For the attached phase, and thus, the vertical branch, we can assume that almost the entire luminosity stems from the accretion of planetesimals onto the core, so that 
\begin{equation}
    \label{eq:Lcore_appB}
    L = L_\mathrm{M,s} = L_\mathrm{core} = \frac{G M_c \dot{M}_c}{R_c}.
\end{equation}
Also, the assumption 
\begin{equation}
    \Mpla = M_\mathrm{core}
\end{equation}
holds relatively well throughout the full attached phase as we can see in \cref{fig:defaultCasePlots}.
The radius in \cref{eq:R_out} reduces to 
\begin{equation}
\label{eq:appB_ReqRacc}
    \Rout = \RBondi
\end{equation}
in the early attached phase and to 
\begin{equation}
\label{eq:appB_ReqRhill}
    R = \frac 14 R_\mathrm{H}
\end{equation}
later on before detaching \citep{lissauer_models_2009}.
We estimate the planetesimal accretion rate $\dot{M}_\mathrm{c}$ of a young protoplanet with a statistical estimation according to \citet{armitage_astrophysics_2020} which depends on the surface density of planetesimals in the disk $\Sigma_\mathrm{pla}$, the Keplerian frequency $\Omega$, the core radius $R_\mathrm{c}$ and the gravitational focusing factor \citep{safronov_formationSolarSystem_1984}
\begin{equation}
\label{eq:FocusingFactor_appendix}
F_\mathrm{g} = 1 + \frac{v_\mathrm{esc}^2}{v_\mathrm{p}^2}
\end{equation}
with
\begin{equation}
\label{eq:appB_Mdotcore}
    \dot{M}_\mathrm{c} = \frac{\sqrt{3}}{2} \Sigma_\mathrm{pla} \Omega \pi R_\mathrm{c}^2 F_\mathrm{g}.
\end{equation}
We are using this result first for a very young protoplanet whose temperature is still close to the midplane temperature of the disk $T_\mathrm{d}$ and the radius approximation from \cref{eq:appB_ReqRacc} still applies, and later for a more massive planet whose radius is better approximated by \cref{eq:appB_ReqRhill} and its temperature has exceeded $T_\mathrm{d}$ so that $T=T_\mathrm{int}$.

\subsection{The early ascending branch}
\label{sec:appB-Part1-earlyAttachedSlope}

We see in \cref{eq:Bondi-radius} that the Bondi radius depends not only on the planet's mass but also on the speed of sound $c_\mathrm{s}$ in the surrounding medium and thus, the disk's midplane temperature $T_\mathrm{d}$. This leads to an increased dependency on $T_\mathrm{d}$ for $L$:
\begin{align}
    L &= 4 \pi \RBondi^2 \sigma \left(T_\mathrm{int}^4 + T_\mathrm{d}^4\right) \\
    \label{eq:appB_lumiForReqRacc}
    &= 16 \pi \frac{G^2 M_\mathrm{c}^2}{\chi^2}\sigma \left( \frac{T_\mathrm{int}^4}{T_\mathrm{d}^2} + T_\mathrm{d}^2 \right),
\end{align}
where, recalling \cref{eq:cs2defchi},
\begin{equation}    
c_\mathrm{s}^2 = \Gamma_1\frac{\kB T_\mathrm{d}}{\mu \amu} \equiv \chi T_\mathrm{d}. 
\end{equation}
In \cref{eq:appB_lumiForReqRacc} we see that there exists no simple $L-T$ relation to determine the slope on the HR diagram, since $L$ still depends on the (cooling) disk. This explains the initial slope in the HR diagram in \cref{fig:defaultCasePlots} panel (a), in the time when $T_\mathrm{int} \approx T_\mathrm{d}$ (see \cref{fig:defaultCasePlots} panel (b) and remember that we show $T$ and $T_\mathrm{d}$ and not actually $T_\mathrm{int}$).

\subsection{The late ascending branch}
\label{sec:appB-Part2-lateAttachedSlope}

In the shear dominated regime we approximate the planetesimal velocity with $v_\mathrm{p} = \Omega R_\mathrm{H}$ so that the focussing factor reduces to 
\begin{equation}
\label{eq:Fg_simplification}
    F_\mathrm{g} \approx \frac{v_\mathrm{esc}^2}{v_\mathrm{p}^2} = \frac{2GM_\mathrm{c}}{R_\mathrm{c}} \frac{1}{\Omega^2 R_\mathrm{H}^2}.
\end{equation}
Using this in \cref{eq:appB_Mdotcore} results in 
\begin{equation}
    \dot{M}_\mathrm{c} = \frac{\sqrt{3}\Sigma_\mathrm{pla}\pi R_\mathrm{c} G M_\mathrm{c}}{\Omega R_\mathrm{H}^2}.
\end{equation}
We use this result to further specify the luminosity from \cref{eq:Lcore_appB} and compare it to the actual luminosity from \cref{eq:LsigmaT4} with the radius $R=\frac 14 R_\mathrm{H}$:
\begin{equation}
    \frac{G M_\mathrm{c}}{R_\mathrm{c}} \frac{\sqrt{3}\Sigma_\mathrm{pla}\pi R_\mathrm{c} G M_\mathrm{c}}{\Omega R_\mathrm{H}^2} 
    =
    \frac{\pi}{4} R_\mathrm{H}^2 \sigma T^4 
\end{equation}
We further assume $T = T_\mathrm{int}$ and neglect the contribution from $T_\mathrm{d}$. This is fairly accurate in this later stage of the attached phase because $T_\mathrm{int} \gg T_\mathrm{d}$ and because of the fourth power dependence in the boundary condition in \cref{eq:tempAttachedPhase}.
With $R_H = a \left ( {M_\mathrm{c}}/{3 M_\star} \right )^{1/3}$, solving for $M_\mathrm{c}$ results in 
\begin{align}
    M_\mathrm{c} &= \left( \frac{\sigma}{4} \frac{\Omega}{\sqrt{3}\Sigma_\mathrm{pla}G^2} a^4 \left( \frac{1}{3M_\star} \right)^{\frac 43} \right)^{3/2} T^6 \\
    &\equiv B T^6.
\end{align}
Now that we know the dependency of the (core) mass on the temperature, we can also conclude on the dependency of the luminosity on the temperature with \begin{align}
    L &= \frac{\pi}{4} \left( \frac{M_\mathrm{c}}{3M_\star}\right)^{2/3} a^2 \sigma T^4 \\
      &= \frac{\pi}{4} \left( \frac{B T^6}{3M_\star}\right)^{2/3} a^2 \sigma T^4 \\
      &=\frac{\pi\sigma^2T^8}{16\!\sqrt{3}\,\Sigma_\mathrm{pla}}\left(\frac{a^3}{GM_\star}\right)^{3/2}.  %
\end{align}
We conclude that for the later stage of the attached phase where $R \approx \frac 14 R_\mathrm{H}$ we expect that $L \propto T^8$, resulting in a steep line on the HR diagram. 

\end{appendix}

\end{document}